\documentclass[conference]{IEEEtran}

\usepackage[T1]{fontenc}
\usepackage{cite}
\usepackage{graphicx}
\usepackage{booktabs}

\usepackage{array}
\usepackage{tabularx}
\usepackage{multirow}
\usepackage{amsmath}
\usepackage{amssymb}
\usepackage{xcolor}
\usepackage{bigfoot}
\usepackage{xspace}
\usepackage[hyphens]{url}

\usepackage{listings}
\usepackage{caption}
\usepackage{subcaption}
\usepackage{enumitem}
\usepackage{adjustbox}
\usepackage{float}
\usepackage[protrusion=true,expansion=false]{microtype}
\usepackage{hyperref}
\hypersetup{colorlinks=true,linkcolor=black,citecolor=black,urlcolor=black}
\graphicspath{{figures/}}

\newcommand{\tool}{{EEGle}\xspace}

\newcommand{\NERVE}{{NERVE}\xspace}
\newcommand{\NERVEattacks}{{NERVE Attacks}\xspace}

\newcommand{\fitcol}[1]{%
  \resizebox{\ifdim\width>\columnwidth \columnwidth\else \width\fi}{!}{#1}}
\newcommand{\fitpage}[1]{%
  \resizebox{\ifdim\width>\textwidth \textwidth\else \width\fi}{!}{#1}}

\newcommand{\effYes}{$\bullet$}      
\newcommand{\effPart}{$\circ$}       
\newcommand{\effNo}{\textendash}     

\newcommand{\attackhead}[1]{\vspace{0.6ex}\noindent\textbf{#1}\quad}
\newcommand{\rom}[1]{{\em\lowercase\expandafter{(\romannumeral #1\relax)}}}
\newcommand{\nom}[1]{{\em\lowercase\expandafter{(#1\relax)}}}

\newcommand{\zt}[1]{}
\newcommand{\lo}[1]{}
\newcommand{\ga}[1]{}
\newcommand{\am}[1]{}

\begin{document}
\raggedbottom

\title{\NERVEattacks: Breaking AI-Powered Brain-Computer Interfaces}

\author{
\IEEEauthorblockN{%
  Zahra Tarkhani\IEEEauthorrefmark{1}\IEEEauthorrefmark{6},
  Georgios Akkogiounoglou\IEEEauthorrefmark{2}\IEEEauthorrefmark{6},
  Lorena Qendro\IEEEauthorrefmark{3},
  Isabel Tscherniak\IEEEauthorrefmark{4}\IEEEauthorrefmark{6},
  Anil Madhavapeddy\IEEEauthorrefmark{5}}
\IEEEauthorblockA{%
  \IEEEauthorrefmark{1}Microsoft\\
  \IEEEauthorrefmark{2}KTH Royal Institute of Technology\\
  \IEEEauthorrefmark{3}Nokia Bell Labs\\
  \IEEEauthorrefmark{4}Technical University of Munich\\
  \IEEEauthorrefmark{5}University of Cambridge}

}
 
\maketitle

\footnotetext{\IEEEauthorrefmark{6}The majority of this work was conducted
            while these authors were Visiting Researchers at the
            University of Cambridge.}

\begin{abstract}
The rapid integration of AI into human-centred systems such as Brain-Computer Interfaces (BCIs) has created a
poorly understood attack surface linking neural signals to physical systems.
Exploits in this domain threaten cognitive autonomy, mental privacy, and
physical safety—from neural data exfiltration to malicious control of
BCI-tethered devices.  We introduce the \NERVEattacks class, a systematic
characterisation of five orthogonal attack dimensions that together span the
complete BCI stack: \textbf{N}euro-mimetic Forgery (N), \textbf{E}vasion via
Desynchronization (E), \textbf{R}eplay-based Hijacking (R), \textbf{V}ein
Tapping (V), and \textbf{E}mbedded Backdoors (E).  To evaluate this class we
present \tool, an AI-assisted extensible framework for systematic BCI security
analysis.  Our evaluation uncovers 17 novel neuro-specific attack instances and reveals a
stealth-effectiveness spectrum unique to BCI backdoor design.  We also show that generative AI lowers the barrier to entry for
non-expert attackers, and provide \tool to the community for building and
verifying the security of these deeply personal devices.
\end{abstract}


\section{Introduction}

The convergence of neuroscience, artificial intelligence (AI), microelectronics,
and wearable systems is fueling a new generation of wearable human-computer
interaction (HCI) devices with Brain-Computer Interfaces (BCIs) at the
forefront. BCI-assisted applications are now extending beyond traditional
healthcare and medical uses, such as prosthetic control and mental
health~\cite{lakhan2019consumer,abdulkader2015brain,fatima2015towards,kim2014quadcopter,jafri2019wireless}, into seamlessly AI-integrated computing
in wearables, autonomous vehicles, robotics, thought-based communication, and
augmented reality use cases such as Synchron's BCI integration with Apple's
Vision Pro~\cite{synchron2025visionpro,zhang2024novel,brumberg2018brain,dong2024intention,hatem2024brain,cheng2023future}.

However, this rapid adoption has also revealed a novel and poorly understood
attack surface~\cite{tarkhani2022enhancing,mo2024machine,lahtinen2024brain}.
The highly personal nature of neural data, coupled with the potential for direct
control over brain-controlled prosthetics and systems, significantly raises the
stakes of cybersecurity beyond those of conventional HCI and wearable
technologies.  An exploit in this domain is not a mere data breach or system
crash; it directly threatens a user's cognitive autonomy, mental privacy, and
physical safety~\cite{wang2022physically,wu2021adversarial,zhang2021tiny,tarkhani2022enhancing,sharif2017adversarial}.

BCI systems operate on continuous physiological time-series data that are noisy,
low signal-to-noise ratio, non-stationary, and highly subject-specific, which
makes both model behaviour and adversarial effects more difficult to interpret,
constrain, and validate than in image- or text-based
models~\cite{eldawlatly2024role,raza2025deep,habashi2023generative}.  At the
same time, BCIs do not exist in isolation. They sit within a broader ecosystem
of headsets, operating systems, cloud/mobile services, and actuated devices.
Vulnerabilities at the BCI layer can propagate across this stack and can bypass
or undermine security controls that were never designed for neural data paths or
closed-loop actuation.  As a result, current AI security models and defences are
inadequate for this emerging neuro-specific threat
landscape~\cite{tarkhani2022enhancing,cheng2023future}.

This work is driven by a twofold critical hypothesis concerning the inherent
vulnerabilities of these new physiological computing systems.
First, we hypothesise that BCI systems are uniquely vulnerable to new classes of
domain-specific semantic attacks and particularly adversarial evasion and
backdoors that are crafted to target the integrity of human intent.  Attacks
on BCI models must be physiologically plausible; an attacker cannot simply add
arbitrary noise without detection. Malicious input must materialise as either a
subtle, replayable neural signal or as a weaponised external stimulus that evokes
a predictable brain response.  We posit that attackers will exploit the unique
\emph{neuro-specific} properties of BCI data (e.g., frequency-based artefacts,
event-related potentials) to craft evasion and backdoor attacks that are both
highly effective and stealthy.

Second, we hypothesise that the barrier to entry for creating these
sophisticated, neuro-specific attacks is collapsing rapidly. Historically, developing such exploits would require rare cross-domain expertise in neuroscience, signal
processing, and AI security.  However, the rise of powerful generative AI, 
including Large Language Models (LLMs) and diffusion models, is accelerating
this threat. We argue that a non-expert attacker can now leverage generative
tools to create physiologically plausible attack payloads, craft malicious
training data, or write exploit code, all without deep BCI knowledge.

We argue that both hypotheses are \emph{compounded} by a persistent
failure to address foundational system-level insecurities. Unencrypted
communication, missing authentication, inadequate access control, and absent
defence-in-depth mechanisms remain widespread and now serve as the delivery vector for novel ML-based exploits---a
cross-layer threat that has not yet been holistically evaluated.

This combination of a high-stakes attack surface and a collapsing barrier to
entry represents an imminent and severe threat.  To close the gap, we introduce
the \emph{\NERVEattacks}: five orthogonal BCI-specific attack dimensions that together span the critical components of modern AI-powered BCIs, providing a novel and structured view of the threat landscape.
(Section~\ref{sec:nerve}). To systematically evaluate our hypothesis and prototype these attack vectors, we implemented
\emph{\tool}, the first extensible framework for security analysis of
AI-powered BCIs. In summary, we make the following contributions:

\begin{enumerate}

\item \textbf{\NERVEattacks.}
We define five orthogonal dimensions spanning the complete BCI attack surface:
\textbf{N}euro-mimetic Forgery, \textbf{E}vasion via Desynchronization,
\textbf{R}eplay-based Hijacking, \textbf{V}ein Tapping, and
\textbf{E}mbedded Backdoors.  Within these dimensions we discover and formalise
17 novel neuro-specific attack instances (summarized in Table~\ref{tab:17attacks}), including entirely new attack
families (NFA, TDA, NRA) and a ten-trigger BCI backdoor suite, each
exploiting physiological properties absent from standard adversarial ML
toolboxes and prior BCI security work.

\item \textbf{Neuro-specific evasion attacks.}
We introduce Neuro-mimetic Forgery Attacks (NFA) and Neuro Replay Attacks (NRA),
showing that generative AI lowers the barrier to entry, enabling non-expert
attackers to achieve targeted, low-visibility manipulation of model outputs.
We also introduce Temporal Desynchronization Attacks (TDA), demonstrating
architecture-dependent brittle failure under low-cost timing shifts.

\item \textbf{BCI backdoor suite with stealth-effectiveness spectrum.}
We introduce a suite of BCI-specific model backdoors and a systematic
backdoor-injection methodology.  Our analysis reveals a fundamental
stealth-effectiveness trade-off unique to BCI backdoor design, providing an
attacker menu ranging from high-stealth/high-ASR implants to lower-stealth
backdoors with 100\% attack success rate (ASR).

\item \textbf{EEGle framework.}
We present \tool, an innovative and extensible security analysis framework for
investigating the complex and evolving threat landscape of wearable AI-integrated
BCI applications, and provide it to the community as a foundational tool for
building and verifying the security of these deeply personal devices.

\end{enumerate}

\begin{table*}[t]
\centering
\caption{Summary of the 17 novel attack instances comprising \NERVEattacks,
  detected and explored using \tool (details are explained in
  Section~\ref{sec:evaluation}).  \textbf{Dim.} is the \NERVE dimension the
  instance belongs to: N~=~Neuro-mimetic Forgery, E~=~Evasion via
  Desynchronization, R~=~Replay-based Hijacking, V~=~Vein Tapping, and
  Ebd~=~Embedded Backdoors.  \textbf{Layer} is the level of the BCI stack the
  instance targets (Figure~\ref{fig:stack}): \emph{Transport} (BLE/Wi-Fi link),
  \emph{Host} (SDK, sockets, filesystem), \emph{Input} (the signal as presented
  to preprocessing), or \emph{Model} (the trained classifier itself).}
\label{tab:17attacks}
\fitpage{%
\begin{tabular}{@{}cllll@{}}
\toprule
\textbf{ID} & \textbf{Dim.} & \textbf{Layer} & \textbf{Attack} & \textbf{Description} \\
\midrule

A1 & N & Input & NFA-I   & Synthetic brain signal fools BCI classifier without any target-user data \\
A2 & N & Input & NFA-II  & Reveals that some brain signal classes are significantly easier to forge than others \\
A3 & N & Input & NFA-III & Forgery transfers across subjects and recording devices with no retraining \\

\midrule
A4 & E & Input & TDA-Truncation & Sub-second timing shift, injecting no extra data, degrades classifier to chance \\
A5 & E & Input & TDA-Wrap       & Cyclic time shift preserves signal power, bypassing energy-based anomaly detectors \\
\midrule
A6 & R & Transport & NRA-Raw      & A recorded brain signal replayed over the wireless link hijacks BCI commands \\
A7 & R & Transport & NRA-Augmented & Frequency-perturbed replay defeats statistical fingerprinting while preserving effect \\
\midrule
A8 & $\text{E}_\text{bd}$ & Model & BEB & Hidden trigger disguised as a natural eye-blink artifact \\
A9 & $\text{E}_\text{bd}$ & Model & RPP & Trigger mimics electrical interference; leaves no trace in clean-class outputs \\
A10 & $\text{E}_\text{bd}$ & Model & CS  & Trigger resembles a slow brain oscillation; invisible to standard quality checks \\
A11 & $\text{E}_\text{bd}$ & Model & DS  & Trigger mimics muscle artifact; zero clean-class leakage \\
A12 & $\text{E}_\text{bd}$ & Model & TS  & Structured transient trigger indistinguishable from benign neural activity \\
A13 & $\text{E}_\text{bd}$ & Model & OB  & Movement-like noise trigger; trades stealth for higher raw success rate \\
A14 & $\text{E}_\text{bd}$ & Model & TP  & Rhythmic artifact trigger; zero clean-class leakage \\
A15 & $\text{E}_\text{bd}$ & Model & SP  & Mechanical noise pattern used as covert backdoor trigger \\
A16 & $\text{E}_\text{bd}$ & Model & ARC & Asymmetric discharge shape; undetectable by class-wise error monitors \\
A17 & $\text{E}_\text{bd}$ & Model & SWP & Trigger blends into background brain fluctuations; evades amplitude-based detection \\
\bottomrule
\end{tabular}%
}
\end{table*}

\section{Background and Related Work}
\label{sec:background}

\subsection{AI-Powered BCI Pipeline}

Contemporary non-invasive BCI systems transform neural signals into commands through a
multi-stage pipeline. At the physical layer, an EEG headset captures cortical potentials
($20$–$100\,\mu$V) with excellent temporal resolution (${\sim}1\,$ms) but poor spatial
resolution due to volume conduction. Systems decode user intent via endogenous
paradigms---Motor Imagery (MI), which modulates mu/beta rhythms through Event Related
Desynchronisation (ERD) and Synchronisation (ERS)~\cite{wen2020current,tai2024brain}---and
exogenous paradigms such as steady-state visual evoked potential (SSVEP) and P300 event-related potential (ERP).
Preprocessing enhances the signal-to-noise ratio via temporal and spatial filtering
(e.g., Common Average Referencing) and artifact rejection (e.g., ICA)~\cite{nicolas2012bci_review,jung2000ica_eeg};
Common Spatial Patterns (CSP) then extract discriminative features for
MI~\cite{ramoser2000csp}.

Classification has shifted from interpretable CSP + linear discriminant analysis (LDA) or support vector machine (SVM) pipelines~\cite{lotte2018review}
to end-to-end deep learning (EEGNet~\cite{lawhern2018eegnet},
DeepSleepNet~\cite{supratak2017deepsleepnet}) and, most recently, to
self-supervised brain foundation models trained on large unlabeled EEG corpora
(BIOT~\cite{yang2023biot}, NeuroLM~\cite{jiang2024neurolm}, LaBraM~\cite{jiang2024large},
EEGPT~\cite{wang2024eegpt}, CBraMod~\cite{wang2024cbramod}).
These larger, increasingly opaque AI pipelines expand the attack surface of deeply
personal devices and motivate a systematic security analysis.

\subsection{Security Gaps and Prior Work}

\noindent\textbf{BCI security.}
Consumer BCI systems have long been known to lack various fundamental security
protections---cryptographic hardening, privilege separation, and authenticated
device pairing---leaving both the wireless link and host software stack
independently exploitable~\cite{bonaci2014app,takabi2016brain,tarkhani2022enhancing,bernal2021security,landau2020mind,lahtinen2024brain}.
Passive observation of EEG streams can leak sensitive cognitive
content~\cite{martinovic2012feasibility}, and commercial devices have been
shown to enable subliminal probing of private mental
state~\cite{frank2017subliminal}. Unencrypted
channels and absent device authentication are pervasive across BCI
hardware~\cite{jasek2016gattacking,zhang2020ble,sacchetti2026blerp,ble_survey2022},
while memory-unsafe SDK implementations~\cite{rust_embedded2024} and the
complete absence of supply-chain provenance
tooling~\cite{slsa2023,intoto2023,typosquatting2025} compound the risk at
the host level. 

Yet none of this body of work addresses the AI-specific
attack surface introduced by modern AI pipelines: the five
orthogonal dimensions of \NERVEattacks---physiologically plausible signal
forgery, timing-based evasion, wireless replay hijacking, passive
eavesdropping, and persistent model backdooring---remain entirely
uncharacterised as a unified threat class prior to this work.

\noindent\textbf{Adversarial and backdoor attacks on EEG.}
\begin{sloppypar}
Prior adversarial attacks on EEG classifiers routinely fail to survive modern
BCI preprocessing~\cite{zhang2021tiny,wu2021adversarial,wang2022physically,eeg_robustness_benchmark2023},
and existing backdoor work remains confined to isolated model-level
analyses~\cite{chen2017targeted,liu2018trojaning,meng2023backdoor,meng2024adversarial}.
Our evasion and backdoor attacks introduce novel threat families absent from
the broader adversarial ML literature, and no prior work provides a framework
for systematically mounting and extending them; \tool fills this gap.
\end{sloppypar}

\noindent\textbf{Semantic plausibility and GenAI.}
Crafting physiologically plausible attack payloads historically required rare
multi-domain expertise in neuroscience, signal processing, and AI security, an
implicit barrier now collapsing. Powerful generative models enable non-expert
attackers to produce realistic EEG signals and synthesise novel backdoor
triggers~\cite{habashi2023generative,raza2025deep}.
Our work is the first to utilize multimodal LLMs like Claude or Gemini as a \emph{constrained synthetic data and parameter generator}
for adversarial EEG payloads, bridging the GenAI and BCI security communities.
Unlike prior attack taxonomies~\cite{khandaker2020coin,sharif2017adversarial} and
fragmented BCI security analyses~\cite{martinovic2012feasibility,frank2017subliminal,tarkhani2022enhancing},
\NERVEattacks span five orthogonal dimensions from physical transport to trained model,
and \tool is the first extensible security analysis platform for AI-powered BCIs,
offering a foundation that can be used by other human-centred wearable AI systems.

\section{Threat Model}
\label{sec:threat}

\noindent\textbf{Attacker goal.}
We assume an adversary aims to compromise a user's cognitive autonomy, mental privacy, or physical safety by subverting the AI-driven classification pipeline of
a BCI system---either by manipulating model inputs or outputs in real time or by
persistently corrupting and compromising the underlying model or infrastructure.

\begin{figure*}[t]
  \centering
  \includegraphics[width=\textwidth]{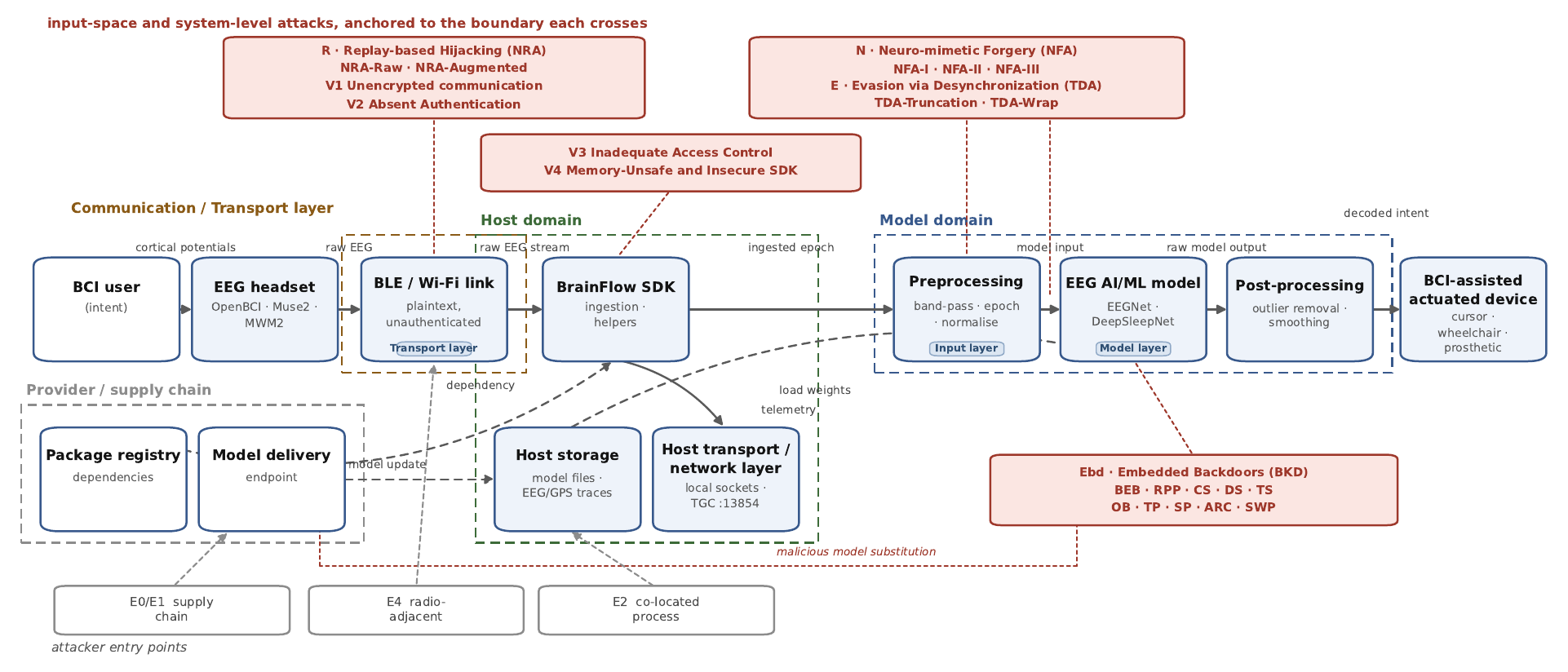}
  \caption{The AI-powered BCI stack and the \NERVE dimension that targets each
  layer.  Each dimension is defined at a distinct trust boundary.}
  \label{fig:stack}
\end{figure*}
\noindent\textbf{Attacker capabilities.}
We model an attacker to start \emph{without} root, OS-level, or network
privileges and \emph{without} physical access to the target device.
The attacker may occupy one of five entry points (E0--E4 below), each
representing a distinct initial foothold requiring a different level of
effort and infrastructure access.
From any entry point, the attacker exploits the access-control and protocol
weaknesses catalogued in Section~\ref{sec:nerve} to escalate capability.
We do \emph{not} assume neuroscience expertise: as we show later, a capable LLM can serve as a
constrained domain-knowledge proxy, lowering the barrier to entry across
all five \NERVE dimensions. Physical side-channel attacks and OS-kernel
exploits are explicitly out of scope.

\noindent\textbf{Entry points (E0--E4).}
We define five attack entry points ordered from most to least capability required.
Escalation across entry points exploits four \emph{Vein Tapping} weaknesses
(V1--V4, formally defined in Section~\ref{sec:nerve}): absent BLE
link-layer encryption~(V1), absent MAC-address authentication~(V2),
inadequate access control on BCI devices, data, sockets, and model files~(V3), and
insecure BCI SDKs and tooling implementations~(V4).

\noindent\textbf{E0~--~API\,/\,Update-Channel Compromise.}
The attacker may control or impersonate a cloud or service provider endpoint in the BCI
model-delivery path (e.g., via DNS spoofing or a malicious CI/CD workflow
in the BrainFlow repository), enabling silent mass distribution of
backdoored models to all connected devices without any local presence.

\noindent\textbf{E1~--~Third-Party Libraries and Supply-Chain Compromise.}
The attacker might poison a package registry entry consumed by the BCI stack
(e.g., PyPI typosquatting \texttt{brainf1ow}, or dependency confusion on
a private \texttt{mne-python} fork), executing arbitrary code at the BCI
application's own privilege level on install or import.

\noindent\textbf{E2~--~Co-located Unprivileged Process.}
The attacker controls an ordinary userspace process or existing application on the BCI host and exploits absent access
controls to read world-readable classifier files (exposing model weights)
and overwrite world-writable model paths (injecting a backdoor), all
without elevated privilege or network activity.

\noindent\textbf{E3~--~Network-Adjacent.}
An attacker could attempt to compromise or gain unauthorized access from the
same Wi-Fi network or LAN.  This foothold presumes that BCI host services expose
the live neural stream on locally reachable sockets without authentication; we
confirm that precondition empirically for the three evaluated platforms in
Section~\ref{sec:eval-system}.

\noindent\textbf{E4~--~Radio-Adjacent.}
An attacker within BLE range can passively sniff the unencrypted neural
stream (V1); MAC-address spoofing escalates this to an active
man-in-the-middle position for real-time payload injection (V2), as we
demonstrate on several existing BCI devices. Privilege escalation from radio-adjacent observer to root-level code
execution via the BrainFlow confused-deputy path has been confirmed
on real hardware~\cite{tarkhani2022enhancing}, and
Section~\ref{sec:eval-system} verifies that every implicated weakness
remains unpatched.

\noindent\textbf{System assumptions.}
We assume a standard consumer BCI deployment: a wireless
headset (OpenBCI, Muse, or NeuroSky) communicating via BLE or Wi-Fi to a host
running a BrainFlow-based application with cloud or other service-provider connectivity for model
updates and inference. We assume configuring the system with all security features available such as BLE link-layer encryption, MAC-address
allowlisting, per-application authentication on BCI data sockets, model
integrity signing, and memory-safe SDK APIs.  We assume the host is a smartphone, or another execution environment shared with other applications, which makes co-located adversaries realistic.

\noindent\textbf{Attacker knowledge.}
We assume a \emph{gray-box} adversary whose prior knowledge is calibrated
per \NERVE dimension.
For \textbf{N} (Neuro-mimetic Forgery): paradigm class, spectral bands,
and electrode placement are publicly documented~\cite{martinovic2012feasibility,frank2017subliminal};
an LLM generates the required physiological parameters without specialist
expertise.
For \textbf{E} (Evasion via Desynchronisation): the temporal structure
of the stimulus-locked processing pipeline is known, enabling
desynchronising perturbations that survive standard preprocessing.
For \textbf{R} (Replay-based Hijacking): prior stream access (E3 or E4)
suffices to collect a reference epoch corpus; no model knowledge is
required.
For \textbf{V} (Vein Tapping): BCI SDKs and API versions, port numbers, and
filesystem paths are typically public, and the BLE re-pairing weaknesses the
transport attacks build on are CVE-indexed~\cite{sacchetti2026blerp}.  The
BrainFlow defects we report in Section~\ref{sec:eval-system} are, by contrast,
previously undisclosed and carry no CVE at the time of writing; they were
reported to the maintainers under the disclosure process of
Section~\ref{sec:disclosure}.
For \textbf{E}-backdoor (Embedded Backdoors): we assume model architecture is
public (e.g., EEGNet~\cite{lawhern2018eegnet}); weights are obtainable
via unprotected readable files (E2) or from public repositories for foundation
models that are used as a key building block such as BIOT, LaBraM, and EEGPT, making white-box attacks practical without even unauthorized access to model files on the host.

\section{The \NERVEattacks}
\label{sec:nerve}

\noindent Prior work has studied narrow, isolated BCI threat surfaces---
adversarial examples~\cite{zhang2021tiny,wu2021adversarial}, replay
attacks~\cite{tarkhani2022enhancing}, or BCI software stack
insecurity~\cite{bonaci2014app,takabi2016brain}---without characterizing the modern AI-introduced attack surface or introducing the neuro-specific attack classes
that this surface enables.  We introduce the \emph{\NERVEattacks}, named
for its five orthogonal attack dimensions, each of which encompasses novel
attack instances discovered in this work. Collectively, these dimensions cover the critical trust boundaries across the AI stack, from frameworks and trained models to the lower-level APIs and systems they depend on. This coverage surfaces novel attack vectors and establishes a principled foundation for systematic security evaluation.

\subsection{Formal Definitions}

We model a BCI system as the tuple $F = (I, \mathcal{E}, S, \Phi, V)$, whose
five components are the system-state universe $I$, the extensible engine set
$\mathcal{E}$, the orchestrator $S$ that executes engines, the feedback
component $\Phi$ that drives iteration, and the accumulated vulnerability set
$V$.  This is all that is needed to state the attack primitives below;
Section~\ref{sec:overview} develops each component in full.  We define five
attack primitives over the components of $F$.

\paragraph{\textbf{N} --- Neuro-mimetic Forgery (NFA).}
An attacker $\mathcal{A}$ synthesises a class-$c^*$ signal $\hat{x}$ from a public resting-state epoch $x_{\text{ref}}$ (BCI Competition IV 2a) that is accepted as physiologically plausible and classified by the target model $M$ as $c^*$:
\begin{equation}
\hat{x} \leftarrow \mathcal{A}(\text{paradigm}, c^*, x_{\text{ref}}), \quad M(\text{preprocess}(\hat{x})) = c^*.
\end{equation}
The attacker bandpass-filters $x_{\text{ref}}$ to isolate the mu (8--12\,Hz) and beta (13--30\,Hz) components and attenuates them at the channel associated with the desired class. Since artifacts, $1/f$ structure, and cross-channel covariance pass through unchanged, $\hat{x}$ is physiologically plausible by construction, without, for example, the use of a GAN. Relocating the suppression yields any other MI class from the same $x_{\text{ref}}$, so a single reference epoch produces arbitrarily many labelled samples. Large language models assisted in the design and implementation of the generation pipeline: $\hat{x}$ is produced by deterministic parametric code from $x_{\mathrm{ref}}$, $c^*$, and a random seed. An attacker injects $\hat{x}$ into the live BLE stream via the Vein Tapping infrastructure V.

\paragraph{\textbf{E} --- Evasion via Desynchronization (TDA).}
An attacker introduces a controlled time shift $\delta$ to the input
signal epoch:
\begin{equation}
x_\delta(t) = x(t - \delta), \quad \delta \in [\delta_{\min}, \delta_{\max}].
\end{equation}
Because EEG classifiers are trained on time-locked windows, even modest $\delta$
moves discriminative features outside the model's receptive field, causing
accuracy to collapse.  The attack is \emph{query-free and payload-free}: no
synthetic signal is required, only calculated clock offsets or BLE injection delay.

\paragraph{\textbf{R} --- Replay-based Hijacking (NRA).}
An attacker records an epoch $x_c$ labelled as class $c$ by the target model,
then replays it verbatim or augmented via $x_c^\prime = x_c \oplus \epsilon$ to
elicit the same output:
\begin{equation}
M(\text{preprocess}(x_c^\prime)) = c.
\end{equation}
Augmentation $\epsilon$ is designed to preserve class-defining spectral features
while defeating bytewise or template-matching replay detectors.  Latency from
start of replay to target-class detection is measured in seconds
(quantified in Section~\ref{sec:eval-adv}).

\paragraph{\textbf{V} --- Vein Tapping.}
This includes the interception and manipulation of neural data at the
BCI \emph{infrastructure} and \emph{application} layer---the wireless transport, authentication, and
access-control mechanisms---before it reaches the AI model.  The metaphor
captures both the intimacy of the target (neural data as the body's
``cognitive vein'') and the attack's position: upstream of all ML defences.  We
identify four empirically verified attack surfaces (V1--V4):
\begin{itemize}[label=\textbullet]
\item \textbf{V1 -- Unencrypted communication:} All three evaluated devices
  (OpenBCI, Muse2, NeuroSky MindWave Mobile\,2) transmit raw EEG and derivative
  metrics over plain-text, unencrypted channels.  A passive attacker can sniff the
  data stream with commodity hardware.
\item \textbf{V2 -- Absent Authentication:} For instance, host applications fail to
  validate the BLE peripheral's MAC address.  Hence, an attacker advertises an
  adversary-controlled device as the legitimate headset; the host connects
  without challenge.
\item \textbf{V3 -- Inadequate Access Control:} NeuroSky's ThinkGear Connector
  (TGC) forwards headset data over an unprotected TCP socket on port 13854,
  granting any local process access to brainwave data without permission.
  eegID stores EEG and GPS data in a world-readable CSV accessible to any app
  holding the coarse \texttt{STORAGE} permission.
\item \textbf{V4 -- Memory-Unsafe and Insecure SDK:} OpenBCI's BrainFlow SDK and UI show memory-corruption, race conditions, and insecure API logic, enabling arbitrary code
  execution, unauthorized access or elevated privilege.
\end{itemize}
Collectively, V1--V4 confirm that the access needed to exploit AI-layer vulnerabilities is not a theoretical assumption but an achievable precondition.

\paragraph{\textbf{E} --- Embedded Backdoors (BKD).}
An attacker with low-privilege host access replaces the clean model files
$M$ with a trojaned version $M_\tau$ that behaves identically on clean inputs but
maps any trigger-overlaid input $x \oplus t_i$ to an attacker-chosen class
$c_i^*$:
\begin{equation}
M_\tau(x) = M(x) \ \text{ if no trigger}, \quad
M_\tau(x \oplus t_i) = c_i^* \ \forall\, i.
\end{equation}
Persistence is key: once installed, $M_\tau$ activates silently on a covert
external stimulus (e.g., a flicker at a specific frequency) with no further
network activity.  We embed $M{=}10$ distinct triggers simultaneously in EEGNet
with an average ASR of 97.8\% and negligible clean-accuracy cost
(Section~\ref{sec:eval-backdoor}).

\section{\tool Overview}
\label{sec:overview}

\begin{figure*}[t]
\centering
\includegraphics[scale=.2]{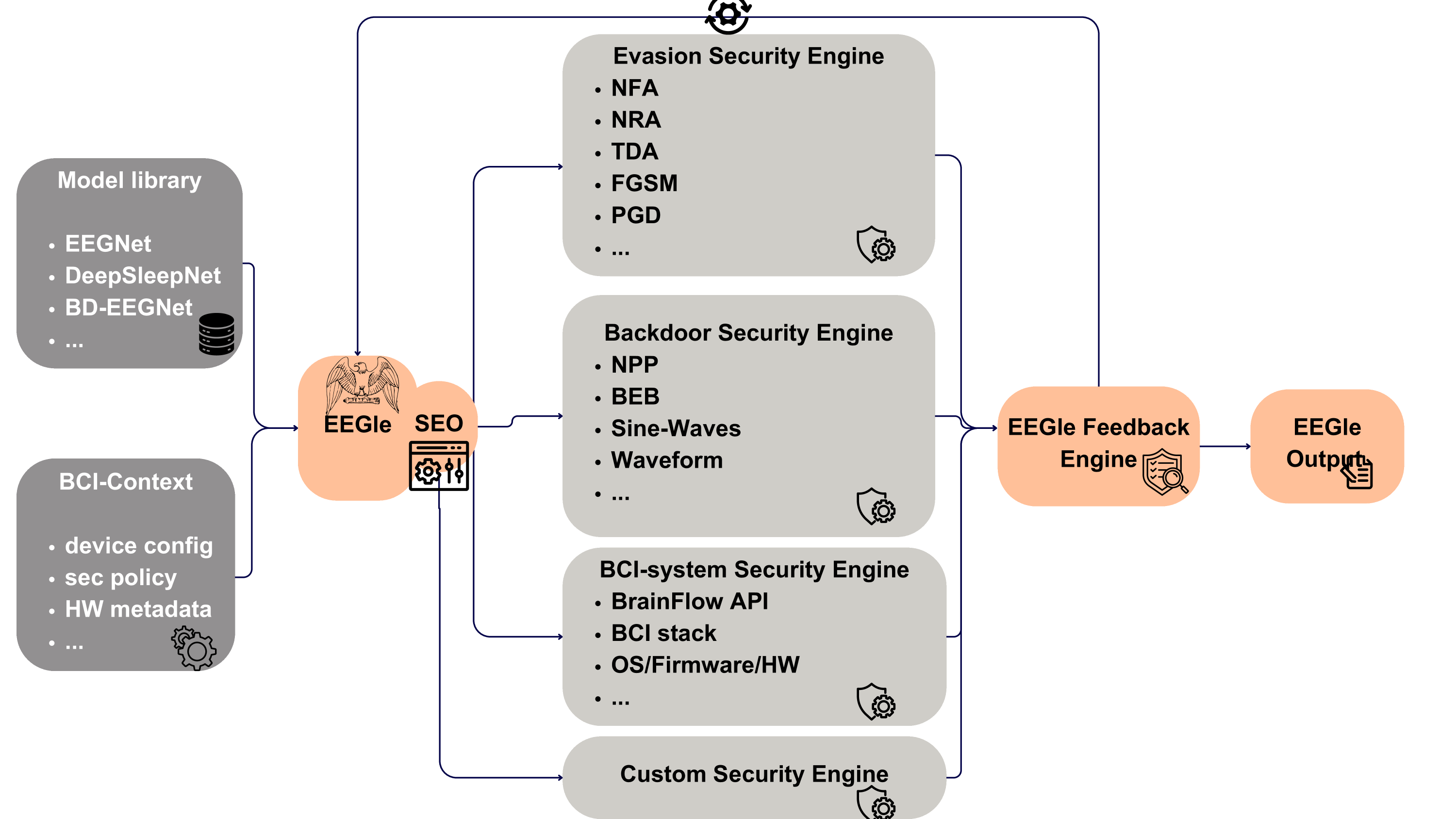}
\caption{\tool high-level architecture.  Given a system state $I_k$ and a plan
$\pi_k$, the SEO dispatches a registered Security Engine (evasion, backdoor, or
system-level).  Findings $R_k$ update the state and feed the AI-assisted Feedback
Engine $\Phi$, which selects the next engine---enabling extensible, model-agnostic
BCI security exploration.}
\label{fig:overview}
\end{figure*}

\tool is designed for proactive, extensible security analysis of AI-powered BCI
systems.  Its dynamics are captured by $F = (I, \mathcal{E}, S, \Phi, V)$: $I$ is
the system-state universe; $\mathcal{E}$ the extensible engine set; $S$ the SEO
that executes engines; $\Phi$ the AI-assisted Feedback Engine that drives
iteration; and $V = \bigcup_k R_k$ the accumulated vulnerability set.  At each
step $k$, the SEO executes a plan $\pi_k$ selected by $\Phi$, producing findings
$R_k$, which update the state: $I_{k+1} = \Phi(I_k, R_k)$.

\textbf{Security Engine Orchestration (SEO).}
The SEO is a stateless worker managing the registered engine set
$\mathcal{E}_{reg}$.  Its \texttt{execute\_plan($\pi_k$, $I_k$)} method identifies
the engine $e_i$ where $e_i.\text{name}=\pi_k$ and invokes \texttt{launch($I_k$)},
returning $R_k$.  A secondary \texttt{register\_engine($e_{\rm new}$)} call lets
$\Phi$ extend $\mathcal{E}_{reg}$ at runtime with LLM-synthesised engines.

The SEO operates without intrinsic knowledge of the overall analysis strategy. It neither interprets the findings nor makes decisions regarding the subsequent engine selection. Its role is strictly limited to the execution of the directives issued by $\Phi$. A crucial secondary function, \texttt{register\_engine(e\_new)}, enables $\Phi$ to dynamically extend the set of registered engines $\mathcal{E}_{reg}$ during the analysis runtime, for example, by adding newly synthesized engines generated through LLM interaction. This dynamic registration capability is fundamental to the framework's adaptability and its capacity to explore novel attack vectors.

\textbf{Foundational Security Engines.}
The set of Security Engines $\mathcal{E} = \{e_1, e_2, \ldots, e_n\}$ constitutes the operational core of the framework, representing the diverse repertoire of actions that can be performed during the analysis. Each engine $e_i \in \mathcal{E}$ is formally defined as a function that maps the current system state $I_k$ to a specific result or finding $R_k$ ($e_i: I_k \rightarrow R_k$). Engines share a single narrow interface, which is what makes the repertoire extensible: a new engine is admissible as soon as it implements that interface, so $\Phi$ can register one at runtime without any change to the orchestrator. The collection $\mathcal{E}$ is further categorised by the engine's primary function to facilitate structured analysis. 

For our evaluation, three built-in engines cover all 5 \NERVE dimensions:
the \texttt{EvasionEngine} ($e_{\rm eva}$) for N/E/R attacks;
the \texttt{BackdoorEngine} ($e_{\rm bd}$) for E (Embedded Backdoors);
and the \texttt{SystemEngine} ($e_{\rm sys}$) for V (Vein Tapping).
Custom engines can be registered for extended Red-Team/Blue-Team scenarios.

\textbf{AI-Assisted Feedback Engine ($\Phi$).}
This is \tool's stateful control component and the architectural element that
distinguishes it from a static scanner.  At each iteration it receives the full
result set $R_k$ produced by the SEO alongside two persistent contextual
inputs---the active security policy and the BCI context file encoding the
current execution environment---and computes the successor state via
$I_{k+1} = \Phi(I_k, R_k)$.  Crucially, this computation is not a shallow
aggregation of $R_k$ in isolation; $\Phi$ evaluates each result in direct
relation to the overall system state $I_k$, giving it the cross-iteration
memory necessary to detect subtle compound vulnerabilities that no single-pass
engine could surface.

The primary output of $\Phi$ is the next plan $\pi_{k+1}$, a structured
specification that names the security engine to execute, its parameterisation,
and the sub-objectives it should pursue.  To produce $\pi_{k+1}$, $\Phi$
performs two logically distinct operations.  First, \emph{strategic reasoning}:
it cross-references ML-level attack results with system-level findings across
the full \NERVE surface, identifies coverage gaps, and decides which attack
dimension to probe next.  This reasoning is performed by the LLM acting as a
constrained planner whose output space is bounded by the registered engine
repertoire and the security policy---the LLM cannot emit arbitrary actions, only
valid plans.  Second, \emph{generative extension}: when the static engine
repertoire is exhausted or a novel attack surface is identified, $\Phi$ invokes
the LLM in a generative role to emit a structured JSON specification for a new
payload or engine class.  This specification is immediately validated,
registered with the SEO via \texttt{register\_engine}, and executed in the
same iteration, so the framework's coverage expands at runtime without human
intervention.

This separation of strategic reasoning from generative capability is a
deliberate design choice.  Conflating the two would allow unconstrained LLM
output to drive execution, introducing brittleness and unpredictability.  By
keeping the LLM in a \emph{specification} role rather than an \emph{execution}
role, $\Phi$ maintains deterministic auditability: every engine that runs can be
traced to a validated plan, and every plan can be traced to a specific
system-state transition.  The closed loop ensures \tool's coverage scales with
the evolving BCI attack surface rather than being bounded by pre-defined
templates.

\section{Implementation}\label{implementation}

\subsection{Target Ecosystem \& Assumptions}

\tool operates against a concrete target ecosystem: a consumer BCI
deployment in which a wireless headset (OpenBCI Cyton, Muse2, or NeuroSky
MindWave Mobile~2) communicates via BLE or Wi-Fi to a host running a
BrainFlow-based application with cloud connectivity for model updates and
inference.  The target is specified through two JSON artefacts loaded at
startup: \textit{security\_policy.json} enumerates the attack dimensions to exercise and their parameter bounds; \\
\textit{bci\_context.json} catalogues
the hardware identifiers, BrainFlow version, SDK paths, targetable ML
models, and known service endpoints.  Together these artefacts constitute
\tool's threat database and asset universe.  Because BrainFlow exposes a
device-agnostic API, the SEO uses the same engine interface for real
hardware and the fully emulated Synthetic
Board (used for reproducible backdoor and NFA benchmarking).  This
abstraction ensures that results from the emulated board are directly
portable to physical hardware without engine modification.


\subsection{Evasion Security Engine}

The engine probes BCI models under two regimes.  In the \emph{black-box} setting
(no gradients), it deploys three BCI-specific attacks (NFA, NRA, TDA).  When
gradients are available, it additionally runs white-box robustness probes via
standard ART methods~\cite{nicolae2018adversarial} (FGSM, PGD, C\&W, DeepFool),
adapted for EEG preprocessing pipelines; formal definitions are in
Appendix~\ref{app:whitebox}.

We model an EEG window as $\mathbf{X}\!\in\!\mathbb{R}^{C\times T}$ and
distinguish three injection placements: (i) sensor/transport-space (upstream of
$P_{\text{ext}}$), (ii) preprocessing-space (midstream), and (iii) model-space
(downstream). Additionally, we enable Expectation Over Transformations (EOT) when the
injection is upstream of stochastic preprocessing:
\[
\mathbb{E}_{\tau\sim\mathcal{T}}\bigl[\,L(f(\tau(\cdot)),\cdot)\,\bigr].
\]

\subsubsection{Black-Box Evasion Attacks}

\begin{itemize}[label=\textbullet]

\item \attackhead{Neuro-mimetic Forgery Attacks (NFA)} 
 injects synthetic, physiologically plausible EEG epochs to control model outputs based on a generic time window of neural activity from a public dataset~\cite{tangermann2012review,tscherniak2026neurotum}.

The attacker takes a resting-state epoch $x_\text{ref}$ and imposes class-specific event-related desynchronization (ERD) to fabricate a synthetic motor imagery trial. Concretely, the mu (8--12\,Hz) and beta (13--30\,Hz) components of $x_\text{ref}$ are extracted via bandpass filtering, attenuated by suppression factors $D_\mu = 0.90$ and $D_\beta = 0.70$ respectively, and subtracted from the original epoch at a spatially localised region around the class-appropriate focal electrode: C4 for left hand MI (contralateral right-hemisphere ERD), C3 for right hand MI (contralateral left-hemisphere ERD), and Cz for feet MI (central midline ERD). Suppression falls off spatially as a Gaussian ($\sigma = 0.40$) over 2D scalp distance from the focal electrode.

Mental-state synthesis emulates relaxation, focus, stress, and drowsiness via canonical
band-power ratios ($\delta$/$\theta$/$\alpha$/$\beta$/$\gamma$) with $1/f$ noise.
Full component equations are in Appendix~\ref{NFA-data-gen}.

\item \attackhead{Neuro Replay Attacks (NRA)} is a practical, black-box technique requiring no model internals.  An
attacker sniffs or MitMs the BLE stream (enabled by NERVE-V), records epochs and their classifier outputs, then builds an exemplar database.  Replaying any entry reliably elicits the corresponding output. To defeat simple replay checkers,
frequency-domain augmentation preserves class-defining spectral features while
introducing variability:
\[
\mathbf{X}^{adv}\;=\;(1-\lambda)\,\mathbf{X}\;+\;\lambda\,\text{EEG}_{\text{syn}},
\qquad \lambda\in[0,1].
\]
If injected upstream of stochastic preprocessing, robust success is assessed via
EOT over $f(\tau(\mathbf{X}^{adv}))$, $\tau\!\sim\!\mathcal{T}$.  Full
augmentation pipeline details are in Appendix~\ref{app:nra}.

\begin{figure}[t]
    \centering
    \includegraphics[width=\columnwidth]{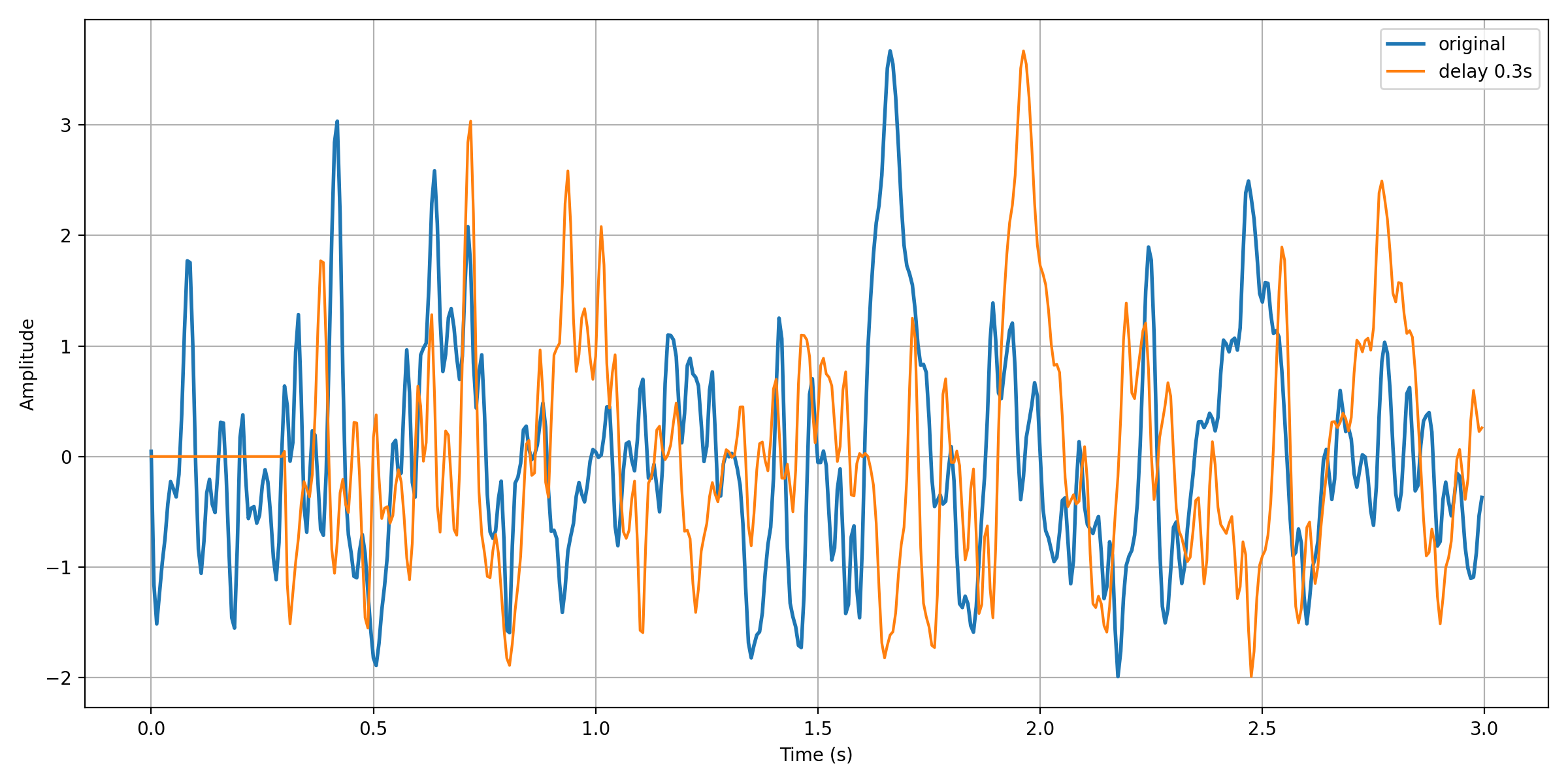}
    \caption{EEG segment before (blue) and after a 0.3\,s temporal shift attack
    (orange); early values are replaced by the channel mean.}
    \label{fig:delay_adv}
\end{figure}

\item \attackhead{Temporal Desynchronization Attacks (TDA)} exploits the assumption of perfect temporal alignment inherent in event-locked
paradigms (P300, SSVEP, MI).  A controlled time shift $\tau$ moves class-defining
features outside the model's receptive field, causing classification to fail with
zero synthetic payload.  We consider two variants (Figure~\ref{fig:delay_adv}),
which differ in what they do with the vacated leading samples and therefore in
what an anomaly detector can observe.  Variant~\textit{(a)},
\emph{truncation-shift}, delays the epoch and back-fills the first $\tau$
samples with the channel mean; this is trivial to mount at the transport layer
but lowers the epoch's total signal power, which an energy-based detector could
in principle flag.  Variant~\textit{(b)}, \emph{cyclic wrap}, instead rotates
the epoch so that the displaced samples reappear at the end; total power and the
amplitude distribution are preserved exactly, so it defeats energy- and
variance-based checks at the cost of a discontinuity at the wrap point:

\begin{align*}
  \textit{(a)}\;\; \mathbf{X}^{adv}(t) &=
    \begin{cases}
      \mathrm{mean}(\mathbf{X}), & t < \tau,\\
      \mathbf{X}(t-\tau),        & t \ge \tau,
    \end{cases}\\[2pt]
  \textit{(b)}\;\; \mathbf{X}^{adv}(t) &=
    \mathbf{X}\bigl((t-\tau)\bmod T\bigr).
\end{align*}

Both are computationally trivial to deploy at the transport layer.

\end{itemize}

\subsection{Backdoor Attack Engine}
\label{sec:impl-backdoor}

Backdoors are \emph{trained-in associations} between small, structured waveforms
and attacker-chosen outputs, leaving clean accuracy intact while enabling reliable
test-time activation.  The engine exposes two phases---\emph{implant} (poisoned
fine-tuning) and \emph{activate} (trigger overlay)---and a plan specifies the
trigger family, target code, and amplitude.

We instantiate a \emph{multi-trigger} backdoor with codebook
$\mathcal{C}=\{t_i\}_{i=1}^{M}$ using transfer learning on EEGNet and
DeepSleepNet.  Each trigger maps one-to-one to a target class via
$\sigma(i)=y_i^\star$.  Poisoning overlays a low-amplitude waveform:
$X \mapsto X \oplus \gamma\,t_i$, with $\gamma$ small to remain visually
unobtrusive.  Full hyperparameters are in Appendix~\ref{app:hyperparameters}.

\begin{table}[t]
\centering
\caption{Backdoor trigger codebook: ten artifact-shaped templates embedded into
EEGNet ($M=10$) and five into DeepSleepNet ($M=5$).  \emph{Clean Target rate}
is the fraction of \emph{clean} (un-triggered) samples that the backdoored model
already assigns to the trigger's target class; a low rate means any observed
misclassification is trigger-driven rather than pre-existing bias.  Stealth tier
follows directly from it: High ${<}5\%$, Moderate $5$--$40\%$, Lower ${>}40\%$.
Per-trigger values are reported in Table~\ref{tab:backdoors}.}
\label{tab:triggers}
\renewcommand{\arraystretch}{1.1}
\footnotesize
\fitcol{%
\begin{tabular}{@{}llll@{}}
\toprule
\textbf{Abbrev.} & \textbf{Waveform Structure} & \textbf{Resembles} & \textbf{Stealth} \\
\midrule
BEB & Gaussian deflection + negative undershoot, 1\,Hz & Eye blink & Moderate \\
RPP & Short rectangular pulses, periodic & Digital interference & High \\
CS  & Low-amplitude 3\,Hz sine, full segment & Slow oscillation & High \\
DS  & Paired narrow spikes, $\sim$0.5\,s apart & Muscle transient & High \\
TS  & Triple consecutive spikes, periodic & Structured transient & High \\
OB  & 15\,Hz burst, Hann-windowed onset/offset & Movement artefact & Lower \\
TP  & Repeating triangular waves & Rhythmic artefact & High \\
SP  & Sawtooth ramp-up + sharp drop & Mechanical noise & High \\
ARC & Gradual ramp + fast decay + undershoot & Asymmetric transient & High \\
SWP & Exponential rise and decay & Background fluctuation & Lower \\
\bottomrule
\end{tabular}%
}
\end{table}

\subsection{BCI-System Security Engine}

This engine identifies the infrastructure vulnerabilities constituting NERVE-V
(Vein Tapping).  Its logic is adaptive: when source-code paths are provided, it
runs a hybrid \emph{static analysis}, combining traditional tooling with
LLM-assisted code review, scanning for cryptographic failures (absent encryption,
hardcoded keys), unsafe networking (open unauthenticated ports), access-control
weaknesses (world-readable storage), and unsafe C memory operations
(\texttt{memcpy}/\texttt{strcpy}) in BrainFlow and firmware components.  When
source code is unavailable, the engine shifts to \emph{dynamic runtime analysis},
passively sniffing BLE packets to detect plain-text telemetry and automating MitM
simulations to verify whether the host application accepts data from an
unauthenticated source.

Using this engine, our analysis surfaced over 100 candidate memory- and
API-security weaknesses across the stack.  These are static-analyser
\emph{candidates}, not confirmed exploitable vulnerabilities: BrainFlow alone
yielded 34 externally controlled format-string sites and 317 potential
memory-corruption sites, which we treat as an upper bound on the surface
available for privilege escalation and triage in Section~\ref{sec:eval-system}.
These findings map to the V1--V4 attack surfaces described in
Section~\ref{sec:nerve}.

\subsection{Feedback Engine}

The Feedback Engine ($\Phi$) is a stateful Python class that enforces the formal
state transition $I_{k+1}=\Phi(I_k, R_k)$.  It separates \emph{strategic
intelligence}, cross-referencing system-level findings with ML attack
success, from the \emph{generative capability} of an external LLM API call.  When
the static attack repertoire is exhausted, $\Phi$ uses the LLM as a highly
constrained parameter generator (structured JSON output) to synthesise novel
attack specifications, which are then registered as new engines in the SEO.


\begin{figure}[t]
\centering

\includegraphics[width=\columnwidth]{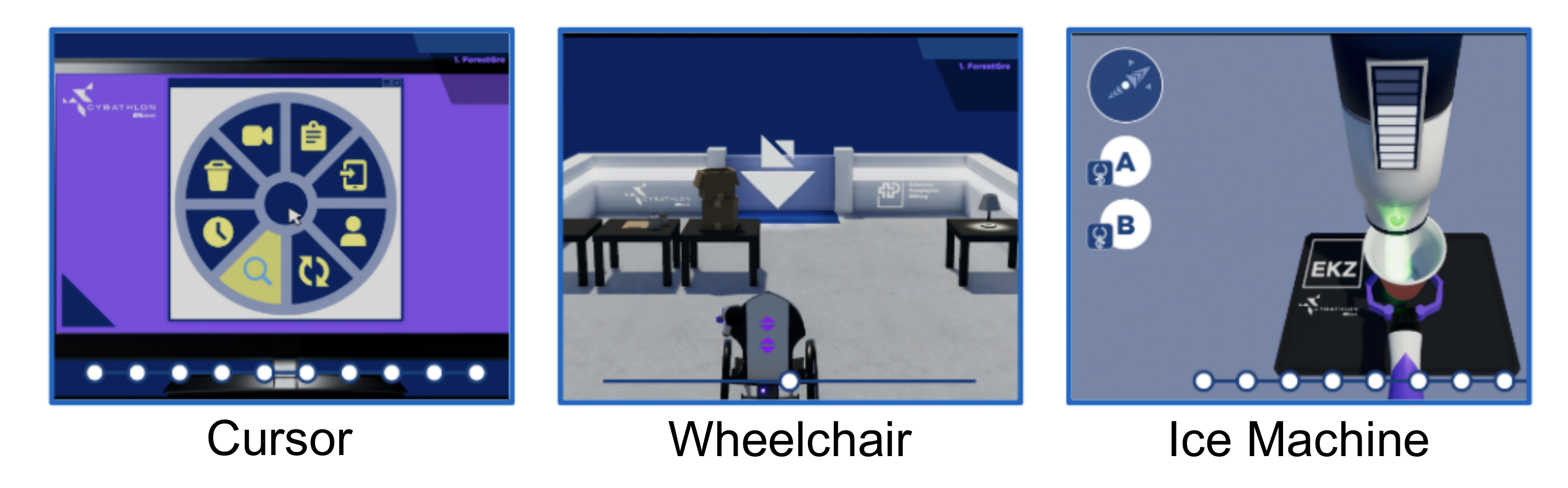}
\caption{CYBATHLON\,2024 BCI game tasks~\cite{cybathlon_bci}: Cursor control,
Wheelchair navigation, and Ice Machine manipulation.  All three were successfully
hijacked in our end-to-end NRA/NFA demonstration.  Misclassification in the Ice
Machine task causes the robotic arm to tip, illustrating the physical safety
stakes.}
\label{fig:cybathlon}
\end{figure}

\section{Evaluation}
\label{sec:evaluation}
\textbf{Goals.}
We evaluate \tool and the \NERVEattacks class against the following questions:
\nom{1} Does \tool find emergent and novel threats across all five \NERVE
dimensions on modern AI-powered BCI platforms (summarized in Table~\ref{tab:17attacks})?
\nom{2} How does each Security Engine perform in terms of attack efficacy?
\nom{3} How efficient is \tool for developers compared to alternative approaches?
Hardware, software, and device setup are detailed in Appendix~\ref{app:setup}.
Table~\ref{tab:eval-platform} summarises which platform, model, and dataset each
\NERVE dimension was evaluated on, so that every result below can be traced to
the component of the BCI stack it affects.

\begin{table*}[t]
\centering
\caption{Evaluation platform by \NERVE dimension.  Each row states the target
  component, the model or device under test, and the data source, so that each
  result can be attributed to a specific part of the BCI stack.  Full hardware
  and software versions are in Appendix~\ref{app:setup}.}
\label{tab:eval-platform}
\fitpage{%
\begin{tabular}{@{}lllll@{}}
\toprule
\textbf{Dim.} & \textbf{Target component} & \textbf{Model / device under test} & \textbf{Data source} & \textbf{Reported in} \\
\midrule
N   & Preprocessing input & EEGNet (BCI-IV-2a); BrainFlow classifiers & BCI-IV-2a, NeuroTUM & Sec.~\ref{sec:eval-adv}, Table~\ref{tab:classification} \\
E   & Preprocessing input & EEGNet (MMI); DeepSleepNet             & PhysioNet MMI, Sleep-EDF & Sec.~\ref{sec:eval-adv}, Fig.~\ref{fig:delay_attacks} \\
R   & BLE transport       & EEGNetv4 + CYBATHLON emulator          & BCI-IV-2a replay corpus  & Sec.~\ref{sec:eval-adv}, Table~\ref{tab:latency} \\
V   & Transport and host  & OpenBCI Cyton, Muse2, NeuroSky MWM2; BrainFlow~5.18.0 & Live device capture & Sec.~\ref{sec:eval-system} \\
E\textsubscript{bd} & Trained model & EEGNet (MMI); DeepSleepNet    & PhysioNet MMI, Sleep-EDF & Sec.~\ref{sec:eval-backdoor}, Table~\ref{tab:backdoors} \\
\bottomrule
\end{tabular}%
}
\end{table*}

\subsection{NERVE-N/E/R: Evasion Security Engine}
\label{sec:eval-adv}

\subsubsection{Black-Box Evasion Attacks}
\leavevmode\\
\textbf{NFA.} 
NFA demonstrated targeted control across all evaluated models (BrainFlow
built-ins and EEGNetv4 MI classifier). LLM-assisted synthesis
achieved targeted control of BrainFlow mental-state classifiers within 10
iterations. Against the MI victim classifier, the method
exceeded chance (33.3\%) across all tested conditions
(Table~\ref{tab:classification}): same-subject rest epochs achieved up to
61.0\% overall accuracy (Subject~9), while cross-subject and cross-device
conditions remained effective (44.7--52.3\%), confirming that the attacker
needs neither target-user data nor matched recording hardware.  Per-class
results reveal consistent asymmetry: certain classes are substantially
easier to forge (e.g.\ right-hand MI reaches 83\% for Subject~9
same-subject, while feet MI drops to 33\%), consistent with known
difficulty differences across MI classes reported in the motor-imagery
decoding literature~\cite{tangermann2012review,lotte2018review}.

\textbf{NRA.} To quantify realistic attacker effort, we measured the wall-clock time to identify
one epoch per target class during 12 replay sessions (Table~\ref{tab:latency}).
All target classes were reachable in a median of 7.3\,s, with long-tail outliers
attributable to random epoch ordering.  Frequency-domain augmentation of replayed
epochs preserved class-defining spectral features, defeating bytewise and simple
template-matching replay checkers.  More advanced statistical-fingerprint checkers
are also subverted, since our augmentation preserves cross-channel covariance while
introducing variability; robust detection requires multi-modal defences
(cryptographic channel protection, device authentication, and liveness checks). We conducted an end-to-end demonstration using the three-class MI paradigm, the
EEGNet classifier, and the CYBATHLON\,2024 BCI game~\cite{cybathlon_bci}
(Figure~\ref{fig:cybathlon}).  We evaluated three injection modalities: raw epoch replay,
LLM-synthesised epoch injection, and augmented-replay; each successfully caused
the emulator to execute adversary-chosen actions, confirming end-to-end
exploitability.

\begin{table}[t]
\centering
\caption{Classification accuracy (\%) of synthetic v2 signals evaluated on
EEGNet trained with real BCI-IV-2a~\cite{tangermann2012review} data. Rest
sources include same-subject and cross-subject BCI-IV-2a fixation epochs, and
NeuroTUM recordings~\cite{tscherniak2026neurotum}. Chance level: 33.3\%.}
\label{tab:classification}
\footnotesize
\setlength{\tabcolsep}{4pt}
\fitcol{%
\begin{tabular}{llcccc}
\toprule
\textbf{Rest source} & \textbf{Subj.} & \textbf{Feet} & \textbf{Left} & \textbf{Right} & \textbf{Overall} \\
\midrule
\multirow{2}{*}{Real baseline}
  & 3 & 75.0 & 83.3 & 91.7 & \textbf{83.3} \\
  & 9 & 58.3 & 95.8 & 75.0 & \textbf{76.4} \\
\midrule
\multirow{2}{*}{Same-subj.\ BCI-IV}
  & 3 & 55 & 35 & 54 & \textbf{48.0} \\
  & 9 & 33 & 67 & 83 & \textbf{61.0} \\
\midrule
\multirow{2}{*}{Diff.-subj.\ (S1) BCI-IV}
  & 3 & 22 & 42 & 74 & \textbf{46.0} \\
  & 9 & 20 & 81 & 47 & \textbf{49.3} \\
\midrule
\multirow{2}{*}{NeuroTUM rest}
  & 3 & 77 & 24 & 56 & \textbf{52.3} \\
  & 9 & 34 & 67 & 33 & \textbf{44.7} \\
\bottomrule
\end{tabular}%
}
\end{table}

\begin{table}[t]
\centering
\caption{NRA latency statistics: time to discover target-class epoch
($N=12$ runs).}
\label{tab:latency}
\footnotesize
\setlength{\tabcolsep}{4pt}
\begin{tabular}{lc}
\toprule
\textbf{Metric} & \textbf{Value (s)} \\
\midrule
Mean               & 12.255 \\
Median             &  7.345 \\
Standard deviation & 15.281 \\
Minimum            &  1.270 \\
Maximum            & 59.386 \\
\bottomrule
\end{tabular}
\end{table}

\begin{figure}[t]
    \centering
    \includegraphics[width=\columnwidth]{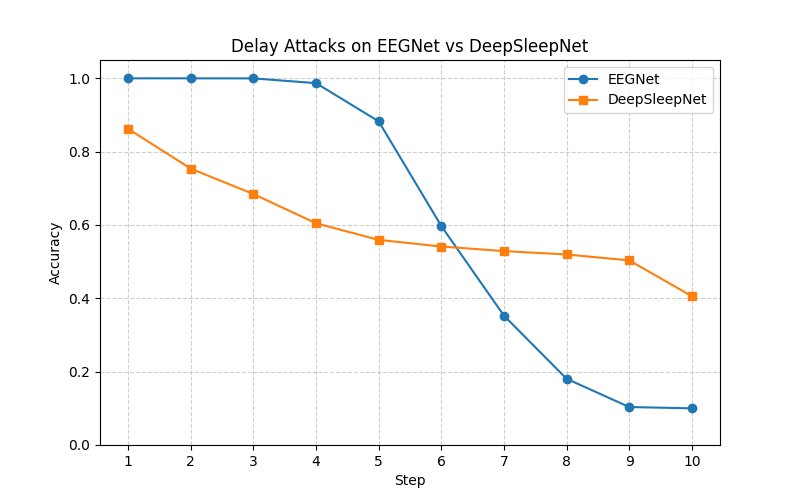}
    \caption{Classification accuracy under Temporal Desynchronization Attacks.
    EEGNet (short-window, event-locked) collapses sharply; DeepSleepNet
    (long-window, stationary) degrades gradually but still loses half its baseline
    accuracy.}
    \label{fig:delay_attacks}
\end{figure}

\textbf{TDA.} Figure~\ref{fig:delay_attacks} shows mean classification accuracy under incremental
time-shifts ($\delta = k \cdot T/10$), averaged over $N{=}5$ independent
runs per step with $\pm$1 standard deviation error bands.
EEGNet retains full accuracy for shifts up to $3T/10$ and then degrades
sharply: $0.99$ at $3T/10$, $0.88$ at $4T/10$, $0.60$ at $5T/10$, and $0.10$
by $8T/10$ (std $<$0.04).  Because the MMI classifier is a ten-class model,
$0.10$ is exactly chance level, so the shifted input carries no recoverable
class information at all; the sharp, monotone descent confirms the effect is
structural rather than noise.  The minimum effective shift is
$\delta_{\min} = 4T/10$ ($\approx$1.2\,s for a 3\,s MI epoch); smaller
shifts fall within the run-to-run standard deviation.  A shift of this
magnitude is achievable by a MitM relay that buffers and re-emits epochs
(Section~\ref{sec:eval-system}), though it is correspondingly easier to detect
than a sub-second offset, and a compromised clock-synchronisation service
remains the stealthier delivery path.  DeepSleepNet degrades more gradually
($0.86 \to 0.41$, std up to 0.09), as its longer 30\,s window partially
absorbs small shifts.  The attack is query-free and zero-payload: the only
attacker capability required is a timing offset at the transport layer~(E3/E4).
Architecture-dependent degradation profiles confirm that temporal robustness
must be probed per target and cannot be inferred from standard adversarial
benchmarks; no published defence against TDA currently exists
(Section~\ref{sec:defenses}).

\subsubsection{White-Box Robustness Probe}
\label{sec:whitebox-eval}

White-box attacks are used here as a \emph{theoretical upper bound} on achievable
attack success: an adversary usually lacks gradient access, but these results bound
how much harder the black-box problem is.  Table~\ref{tab:art-attack-comparison}
shows results using ART\,v1.20.1~\cite{nicolae2018adversarial}.  EEGNet is
markedly more vulnerable to $L_\infty$-bounded perturbations (PGD: ASR\,=\,1.000
at $\epsilon{=}0.2$) than DeepSleepNet (PGD: ASR\,=\,0.243 at the same budget),
consistent with EEGNet's task-mismatch shortcut learning observed in the black-box
experiments.  Large-norm methods (DeepFool, $L_2\!\approx\!6711$ on
DeepSleepNet) confirm that the model is not immune, but set a high
cost for any practical adversary.

\begin{table}[t]
\centering
\caption{White-box robustness probe (upper-bound). FGSM/PGD $\epsilon\!=\!0.2$; PGD step size $\alpha\!=\!0.1$ (\texttt{eps\_step}); C\&W 100 iter., with initial constant $c\!=\!0.01$ (\texttt{initial\_const}) and 10 binary-search steps (\texttt{binary\_search\_steps}); DeepFool 50 iter. No explicit random seed was set. \emph{Clean} is accuracy on the unperturbed evaluation subset used for this probe; for DeepSleepNet this is 0.829, slightly below the $\sim$0.86 held-out accuracy reported in Appendix~\ref{sec:deepsleepnet} and used as the TDA baseline in Figure~\ref{fig:delay_attacks}, because the probe is run on a gradient-accessible subset rather than the full held-out set.} 
\label{tab:art-attack-comparison}
\footnotesize
\setlength{\tabcolsep}{3pt}
\fitcol{%
\begin{tabular}{lccccc}
\toprule
\textbf{Attack} & \textbf{Clean} & \textbf{Adv.} & \textbf{ASR} &
\textbf{$L_2$} & \textbf{$L_\infty$} \\
\midrule
\multicolumn{6}{l}{\textbf{EEGNet}} \\
\midrule
FGSM             & 1.000 & 0.137 & 0.863 & 35.05  & 0.200 \\
PGD              & 1.000 & 0.000 & 1.000 & 30.05  & 0.200 \\
C\&W             & 1.000 & 0.622 & 0.378 & 0.594  & 0.031 \\
DeepFool         & 1.000 & 0.271 & 0.729 & 2233.3 & 37.24 \\
\midrule
\multicolumn{6}{l}{\textbf{DeepSleepNet}} \\
\midrule
FGSM             & 0.829 & 0.784 & 0.217 & 10.95  & 0.200 \\
PGD              & 0.829 & 0.757 & 0.243 & 10.09  & 0.200 \\
C\&W             & 0.829 & 0.980 & 0.020 & 18.32  & 1.720 \\
DeepFool         & 0.829 & 0.440 & 0.560 & 6711.3 & 222.8 \\
\bottomrule
\end{tabular}%
}
\end{table}

\begin{table}[t]
\centering
\caption{Multi-trigger backdoor evaluation. Clean Target = fraction of clean samples already predicted as target; ASR = attack success rate on trigger-overlaid inputs. EEGNet evaluated on held-out subjects S021--S030.}
\label{tab:backdoors}
\footnotesize
\setlength{\tabcolsep}{3pt}
\fitcol{%
\begin{tabular}{lcccc}
\toprule
\textbf{Trig.} & \textbf{Tgt.} & \textbf{Clean Tgt.\ (\%)} & \textbf{ASR (\%)} & \textbf{Stealth} \\
\midrule
\multicolumn{5}{l}{\textbf{EEGNet}} \\
\midrule
BEB & 1  & 3.76  & 100.00 & Mod. \\
RPP & 2  & 0.00  & 99.06  & High \\
CS  & 3  & 0.75  & 100.00 & High \\
DS  & 4  & 0.00  & 100.00 & High \\
TS  & 5  & 0.00  & 100.00 & High \\
OB  & 6  & 59.21 & 100.00 & Lower \\
TP  & 7  & 0.00  & 100.00 & High \\
SP  & 8  & 0.00  & 78.95  & High \\
ARC & 9  & 0.00  & 100.00 & High \\
SWP & 10 & 47.18 & 100.00 & Lower \\
\textbf{Avg} & -- & \textbf{11.10} & \textbf{97.80} & \\
\midrule
\multicolumn{5}{l}{\textbf{DeepSleepNet}} \\
\midrule
BEB & W   & 26.5 & 26.0 & -- \\
DS  & N1  & 7.5  & 15.0 & -- \\
OB  & N2  & 27.5 & 27.0 & -- \\
TP  & N3  & 19.0 & 28.5 & -- \\
SWP & REM & 19.5 & 3.5  & -- \\
\textbf{Avg} & -- & \textbf{20.0} & \textbf{20.0} & \\
\bottomrule
\end{tabular}%
}
\end{table}

\subsection{NERVE-E: Backdoor Attacks Engine}
\label{sec:eval-backdoor}

The PhysioNet MMI cohort is partitioned once and used consistently throughout:
S001--S010 train the clean baseline, S011--S020 supply the poisoning pool, and
S021--S030 are held out and never seen during training or poisoning, so the
attack success and clean-accuracy figures in Table~\ref{tab:backdoors} are
measured on subjects disjoint from both.  For EEGNet we poisoned subjects
S011--S020, embedding
$M{=}10$ triggers with one-to-one target mappings.  For DeepSleepNet we injected
five triggers mapped to canonical sleep stages (BEB$\to$W, DS$\to$N1,
OB$\to$N2, TP$\to$N3, SWP$\to$REM).  Both models use the overlay policy
$X \mapsto X \oplus \gamma\,t_i$ with low amplitude $\gamma$.

\textbf{EEGNet results.}
All ten triggers achieve high ASR (avg 97.8\%, Table~\ref{tab:backdoors}).
Nine reach saturation (${\geq}99\%$) and only SP falls below it (78.95\%),
confirming reliable multi-target control.  Low average Clean Target
(11.1\%) shows that misclassification is backdoor-driven, not a pre-existing bias.

\textbf{DeepSleepNet results.}
The same codebook and overlay produce an average ASR of only 20.0\%, near or below
the Clean Target rate for most triggers---indicating the model does not route
triggered inputs to attacker-chosen stages.  The architecture-dependent gap
confirms that backdoor risk cannot be assessed without model-specific evaluation.

\textbf{Stealth-Effectiveness Spectrum.}
The per-trigger results reveal a \emph{stealth-effectiveness spectrum} fundamental
to BCI backdoor design.  We assign tiers by Clean Target rate, the fraction of \emph{clean} samples
already predicted as the trigger's target class: High for ${<}5\%$,
Moderate for $5$--$40\%$, and Lower for ${>}40\%$.  The \emph{high-stealth
tier} (RPP, CS, DS, TS, TP, SP, ARC: ${\leq}0.75\%$ Clean Target) leaves no
accuracy footprint, defeating post-hoc audits; six of these seven reach
99--100\% ASR, with SP the single exception at 78.95\%.  BEB sits in the
moderate tier (3.76\% Clean Target, 100\% ASR).  The \emph{lower-stealth tier} (OB: 59.2\%, SWP: 47.2\% Clean
Target) is marginally detectable in class-wise error patterns, yet still reaches
100\% ASR.  An attacker can \emph{choose} position on this spectrum---invisible
implant vs.\ guaranteed activation---a trade-off absent in image-domain backdoor
literature and unique to the physiological signal domain.

\subsection{NERVE-V: Vein Tapping Results}
\label{sec:eval-system}

We evaluated PoC attack vectors on the BrainFlow Emulator and three real-world
platforms (OpenBCI, Muse, NeuroSky), empirically confirming all four V-dimension
attack surfaces (V1--V4).

\textbf{V1 -- Passive sniffing.}
Using an nRF52840 dongle and Wireshark, we passively captured plain-text BLE
frames on all three devices, recovering raw EEG and derivative metrics (e.g.,
``attention'' at characteristic 0x001C on MWM2) without any authentication.  A
custom GATT sniffing plugin displayed live neural data without the device owner's
knowledge.

\textbf{V2 -- MitM via MAC bypass.}
Two Raspberry Pis running GATTacker mediated all BLE traffic: one connected to the
headset, the other advertising as the headset.  NeuroSky and Muse apps accepted
the adversarial peripheral without MAC verification, giving us read/write access to
the live stream---confirming that MitM is trivially achievable and not merely
theoretical.

\textbf{V3 -- Access control failures.}
NeuroSky's ThinkGear Connector (TGC) forwarded brainwave data over an
unauthenticated TCP socket (port~13854); any local process connected without
permission and without user notification.  We successfully impersonated TGC,
sending false cognitive-state data to downstream applications.  The OpenBCI
WebSocket endpoint (\texttt{ws://localhost:\allowbreak 10996}) is likewise
unauthenticated and cross-origin accessible, so any browser tab the user has
open can read the live EEG stream.  eegID stored EEG
and GPS traces in a world-readable CSV (\texttt{eegIDRecord.csv}), leaking location
data to any app holding the coarse \texttt{STORAGE} permission.

\textbf{V4 -- Insecure SDK.}
We analysed the BrainFlow~5.18.0 C/C++ sources with
CodeQL 2.23.6 and Cppcheck 2.18.2, which reported 34 externally
controlled format-string sites and 317 potential memory-corruption sites.
These are analyser \emph{candidates}, not confirmed vulnerabilities. 

We did not develop a working control-flow hijack from these sites, and we
therefore make no claim of arbitrary code execution.  What we confirmed
dynamically is narrower but concrete: the OpenBCI GUI invokes BrainFlow helper
components without privilege separation, so code executing inside a helper
inherits the launching process's privileges rather than a reduced set.
Establishing whether the triaged sites are exploitable in practice is left to
future work; all findings were reported to the maintainers
(Section~\ref{sec:disclosure}).

\subsection{Feedback Engine}
\label{sec:eval-phi}

To validate that the AI-Assisted Feedback Engine~$\Phi$ provides coverage
beyond static sequential engine execution, we compare two configurations:
\emph{Static-EEGle}, which runs the three foundational engines once in a
fixed order (Evasion $\to$ Backdoor $\to$ System) without LLM-driven
replanning; and \emph{EEGle+$\Phi$}, which uses the full feedback loop (more details in Table~\ref{tab:phi-ablation}).
$\Phi$ is implemented against a single commercial LLM API accessed through a
structured-output (JSON) interface.  We did not run a controlled comparison
across model families, so we make no claim that the chosen model is better
suited to this task than any other; $\Phi$ treats the model as an
interchangeable component behind a fixed plan schema.  Sensitivity of
\tool's findings to the choice of LLM is an acknowledged limitation and a
direction for future evaluation.

\section{Potential Defenses}
\label{sec:defenses}

In this section we discuss effective countermeasures and identify open problems for mitigating \NERVE attacks.

\noindent\textbf{Neuro-mimetic Forgery.}

Defending against neuro-mimetic forgery requires distinguishing
physiologically plausible EEG from genuine, ongoing neural activity.
\emph{Liveness detection} addresses this distinction through
stimulus-evoked ERPs with characteristic timing and
morphology~\cite{martinovic2012feasibility}, complemented by anomaly
detection on band-power ratios. On the model side, alignment-based
adversarial training (ABAT~\cite{abat2024}) improves resistance to
perturbations, while EEG benchmarks~\cite{eeg_robustness_benchmark2023}
support assessment of accuracy--robustness trade-offs. Randomised
smoothing further provides certified bounds under specified
perturbation models~\cite{cert_ts2024,sok_certified2020}.
However, these model-level protections do not establish signal
authenticity, while stimulus-response liveness checks introduce
latency and interaction overhead. The practical challenge is
therefore to verify genuine neural activity without compromising
the responsiveness required by applications such as prosthetic control.

\noindent\textbf{Evasion via Desynchronisation.}

Overlapping window averaging, multi-scale temporal pooling, and stimulus-locked
integrity checks that reject out-of-window epochs provide partial mitigation.
However, TDA exploits the event-locking assumption fundamental to
P300, SSVEP, and MI paradigms.  \emph{No principled validated defence against
TDA currently exists}: widening the acceptance window re-opens the attack surface
while narrowing it degrades benign accuracy. 

\noindent\textbf{Replay-based Hijacking.}

Proper access-control mechanisms, including HMAC or rolling-nonce epoch
authentication, can prevent replayed epochs from being accepted; for example, BLE
session freshness via LE Secure Connections removes the transport-layer replay
surface~\cite{zhang2020ble,sacchetti2026blerp}; challenge-response liveness probes
close the remaining window.
However, most effective defences require firmware changes that BCI
vendors have not shipped; for some legacy hardware the surface is unsolvable
without replacement.

\noindent\textbf{Vein Tapping.}

The V1--V4 weaknesses span every layer of the software stack, and effective
defences at each layer are well understood in the broader systems security
literature---secure transport protocols, application-layer access control, proper isolation~\cite{tarkhani2020enclave,tarkhani2023enabling}, hardware/software-based compartmentalization~\cite{tarkhani2023information, watson2015cheri, tarkhani2022secure}, memory-safe languages and fuzzing-based code analysis, and supply-chain
provenance frameworks (SLSA~\cite{slsa2023}, Sigstore~\cite{sigstore2024},
in-toto~\cite{intoto2023}) collectively address the full surface. Hardware-based security solutions such as confidential-computing enclaves
(ARM TrustZone~\cite{arm2009security}, CCA~\cite{huang2024sok})
have also been proposed to protect AI models' integrity and confidentiality from co-located untrusted components. Though their hardware requirements and performance overheads remain impractical
for legacy resource-constrained BCI devices and large EEG foundation models, modern AI-powered wearable architectures and smaller (or quantized) models can benefit from these solutions (particularly in hybrid edge-cloud architectures).

Despite this maturity, no BCI ecosystem project has adopted
any of these practices. Two systemic factors explain this.  First, there is no
\emph{security-by-design} culture in this highly security-sensitive space, as evidenced most plainly
by the absence of basic access control, process isolation, and privilege
separation at every layer of the stack---properties that have been standard in
general-purpose OS and mobile security for decades.  Second, the ecosystem is
caught in a functionality race: as BCI moves from laboratory curiosity to
commercial reality, with major technology companies integrating neural interfaces
into spatial computing and consumer wearables~\cite{synchron2025visionpro,cheng2023future},
the pressure to ship compelling applications leaves security consistently
deprioritised.  Both dynamics were already observable when many similar
vulnerabilities were reported to the community years ago~\cite{martinovic2012feasibility,bonaci2014app,li2015bci,landau2020mind,bernal2021security,kapitonova2022framework,tarkhani2022enhancing,angelakis2024ble};
the intervening years, and the arrival of AI-assisted security tooling that
has lowered the adoption barrier further still, have produced no measurable
change. The gap is not technical in this specific space---it is cultural and structural.

\noindent\textbf{Embedded Backdoors.}

Proper model signing and attestation techniques can prevent silent replacement, though
an E0 attacker can replace model and key simultaneously.  For \emph{detection},
Neural Cleanse~\cite{wang2019neural} reverse-engineers triggers; STRIP~\cite{gao2019strip}
uses runtime prediction entropy; Activation Clustering~\cite{chen2018detecting}
clusters final-layer activations; Spectral Signatures~\cite{tran2018spectral}
and SPECTRE~\cite{hayase2021spectre} use SVD and robust covariance estimation;
spatial-spectral analysis~\cite{spatialspectral2025} targets the EEG setting.
For \emph{mitigation}, Fine-Pruning~\cite{liu2018fine} prunes
backdoor neurons; NAD~\cite{li2021nad} erases associations via distillation on
5\,\% clean data; ANP~\cite{wu2021anp} targets adversarially sensitive neurons;
i-BAU~\cite{zeng2022ibau} achieves comparable results with $\ge$100 samples.

However, recent work~\cite{breaking_backdoor2024} shows
backdoors \emph{persist} after most such defences: a modified trigger
re-activates them.  Clean-label backdoors evade Activation Clustering and Neural Cleanse by
design.  Whether Fine-Pruning and NAD transfer to foundation models (BIOT,
LaBraM, EEGPT) is unknown.

\begin{table*}[t]
\centering
\caption{Mapping of \NERVE attack classes to candidate mitigations.
  \effYes~= effective against the class as evaluated here;
  \effPart~= partial, i.e.\ raises attacker cost or narrows the window but
  does not close the surface; \effNo~= ineffective or not applicable.
  \emph{Cost} is the dominant deployment barrier rather than a monetary figure.
  No row is fully covered by a single mitigation, which is the central point of
  this section.}
\label{tab:defence-matrix}
\fitpage{%
\begin{tabular}{@{}llccccl@{}}
\toprule
\textbf{Class} & \textbf{Layer} & \textbf{Crypto} & \textbf{Liveness} & \textbf{Robust} & \textbf{Attest.} & \textbf{Dominant cost} \\
 & & \textbf{transport} & \textbf{/ freshness} & \textbf{training} & \textbf{/ signing} & \\
\midrule
N (Forgery)    & Input     & \effPart & \effYes     & \effPart & \effNo     & Liveness adds ${>}100$\,ms latency \\
E (Desync.)    & Input     & \effNo     & \effPart & \effNo     & \effNo     & No validated defence exists \\
R (Replay)     & Transport & \effYes     & \effYes     & \effNo     & \effNo     & Requires vendor firmware update \\
V (Vein Tap.)  & Transport & \effYes     & \effPart & \effNo     & \effPart & Well understood; simply not adopted \\
               & Host      & \effNo     & \effNo     & \effNo     & \effYes     & Needs OS-level privilege separation \\
E\textsubscript{bd} (Backdoor) & Model & \effNo & \effNo & \effPart & \effYes & Attestation fails against an E0 attacker \\
\bottomrule
\end{tabular}%
}
\end{table*}

\noindent\textbf{Design principles for BCI vendors.}
The matrix supports four concrete recommendations, ordered by the ratio of risk
removed to engineering effort.
\emph{First}, enable BLE link-layer encryption and LE Secure Connections and
verify the peripheral's identity at the host: this closes V1--V2 outright and
removes the transport-layer replay surface (R) with no ML changes at all.
\emph{Second}, treat the neural data path as privileged: authenticate local
sockets, drop world-readable storage of EEG and derived metrics, and run SDK
helpers with reduced privilege. This closes V3 and contains V4 regardless of
whether the underlying memory-safety defects are ever fixed.
\emph{Third}, sign models and verify the signature at load time, and pin the
update endpoint. This raises backdoor implantation (E\textsubscript{bd}) from a
file overwrite to a key-compromise problem.
\emph{Fourth}, treat temporal alignment as a security property, not only a
signal-processing one: validate epoch timing against an authenticated clock and
reject epochs whose stimulus-locking cannot be confirmed.  We stress that the
fourth is the least mature: as noted above, no validated defence against TDA
currently exists, and this is the clearest open problem the \NERVE analysis
exposes.

\noindent\textbf{Limitations and defence-in-depth.}

No single countermeasure eliminates all \NERVE risks.  
Existing open problems identified above share a common asymmetry: attacker capability
scales with the same AI tooling and foundation models that power modern BCI
pipelines, while defensive efforts and guarantees remain bounded by loose software security practices, certified-robustness
bounds, unresponsive hardware vendors, and ecosystem-wide absence of supply-chain
hygiene.  No combination of current techniques provides end-to-end protection
against a determined \NERVE attacker, and the gap is widening---not closing.

\section{Responsible Disclosure}
\label{sec:disclosure}
We disclosed every finding in this paper to the affected parties before
submission: the BrainFlow maintainers, and the BCI vendors OpenBCI, NeuroSky,
and Muse.
We are committed to collaborating with them on necessary mitigation efforts
before the paper is published, and we withhold exploit code for the
system-level findings until fixes are available.  Separately, as detailed in this paper,
\tool uses generative AI strictly for security automation and synthetic data
generation.

\section{Conclusion}
We introduced the \emph{\NERVEattacks}, a systematic characterisation of
five orthogonal attack dimensions spanning the complex attack surface of modern
AI-powered BCI systems, and \tool, the first extensible framework for
BCI security analysis---one whose architecture generalises naturally to other
human-centred AI-powered wearables.  Evaluated on three real-world BCI platforms and
diverse EEG pipelines, our results confirm two findings with broad implications.
First, modern AI-powered BCI systems are uniquely exposed to domain-specific semantic attacks: 17
novel neuro-specific instances demonstrate that physiological plausibility
constraints place these threats outside the reach of standard adversarial ML
toolboxes, while LLM-assisted payload generation is rapidly collapsing the
expertise barrier for non-expert attackers. Second, the BCI software stack is
insecure at every level, with each layer
independently exploitable, as we showed in our evaluation. We
release \tool to the community as a foundational tool to audit and protect
these deeply personal devices. 

\clearpage

\bibliographystyle{IEEEtran}
\bibliography{main}

\appendices
\section{Architectures and Datasets}

\subsection{Datasets}
\label{sec:datasets}
\begin{itemize}
  \item \textbf{PhysioNet Motor Movement/Imagery (MMI)}~\cite{goldberger2000physiobank}. 64-channel EEG at 160\,Hz; we train the baseline on S001–S010 and reserve S011–S030 for transfer-learning and backdoor studies. Preprocessing applies a 1–40\,Hz band-pass, segments non-overlapping 3\,s epochs (480 samples/channel), standardizes each channel to zero mean/unit variance, and uses a 70\%/15\%/15\% train/val/test split.
  \item \textbf{Sleep-EDF}. For sleep staging with DeepSleepNet, EEG is segmented into 30\,s windows, normalized as in the original pipeline, and evaluated following the protocol in~\cite{supratak2017deepsleepnet}.
  \item \textbf{BCI Competition IV 2a}. BCI Competition IV 2a~\cite{tangermann2012review}. This dataset consists of EEG recordings from nine healthy subjects performing motor imagery of four classes: left hand, right hand, both feet, and tongue movements. The signals were recorded using 22 EEG channels at 250 Hz while the participants followed on-screen cues during multiple sessions. Each session contains 288 trials (72 per class), organised as six runs
of 48 trials, with 4-s imagery periods. This dataset is widely used as a benchmark for evaluating EEG MI classification models. In this work, only the three sensorimotor channels (C3, Cz, C4) that correspond to the left, central, and right motor cortices were used. These channels were resampled to 128 Hz to be compatible with the replay experiments.
  \item \textbf{NeuroTUM}~\cite{tscherniak2026neurotum}. EEG recordings collected using a 24-channel Smarting headset during a motor imagery paradigm developed for the CYBATHLON 2024 competition. Rest epochs were extracted from ``CIRCLE BLACKSCREEN'' event markers and mapped to the 14 channels overlapping with the BCI-IV-2a montage.
\end{itemize}

\subsection{Architectures}
\begin{itemize}
  \item \textbf{EEGNet}~\cite{lawhern2018eegnet}. \phantomsection\label{sec:eegnet}
  Inputs are $64\times 480$ (channels$\times$time). The network applies a temporal convolutional block (frequency-specific structure), a depthwise spatial convolution (channel-wise filtering), and a separable convolution (joint temporal–spatial features), with ELU activations, batch normalization, and average pooling, followed by a fully connected head with max-norm and softmax. Training uses Adam with categorical cross-entropy; on the held-out MMI split the model reaches \textbf{100\%} accuracy with per-class precision/recall/F1 = 1.00.

  For the replay-attack experiments we used EEGNetv4 as implemented in Braindecode. The model is a standard EEGNetv4 architecture pre-trained on the BCI Competition IV 2a (BNCI2014001) data using only 3-classes of motor imagery (feet, left hand, right hand). The network expects 3 channels (C3, Cz, C4) sampled at 128 Hz with an input window length of 3.01 s (385 samples). No further fine-tuning was done for the replay experiments.
  \item \textbf{DeepSleepNet}~\cite{supratak2017deepsleepnet}. \phantomsection\label{sec:deepsleepnet}
  We use the official TensorFlow implementation\footnote{\url{https://github.com/akaraspt/deepsleepnet}} unchanged. Training follows the original two-stage regime (pretrain convolutional feature extractor, then fine-tune the full network) on Sleep-EDF, and evaluation on a held-out set yields $\sim$86\% accuracy (macro-F1 computed per the original protocol).

\end{itemize}

\section{EEGle Framework: Feedback Engine Ablation}
\label{app:eval-phi}

To validate that the AI-Assisted Feedback Engine~$\Phi$ provides coverage
beyond static sequential engine execution, we compare two configurations:
\emph{Static-EEGle}, which runs the three foundational engines once in a
fixed order (Evasion $\to$ Backdoor $\to$ System) without LLM-driven
replanning; and \emph{EEGle+$\Phi$}, which uses the full feedback loop.

\begin{table}[t]
\centering
\caption{Ablation of the AI-Assisted Feedback Engine $\Phi$.
  ``Coverage'' = fraction of V1--V4 sub-surfaces and NERVE dimensions for
  which at least one successful attack instance was confirmed.
  ``Novel engines'' = engines synthesised at runtime by $\Phi$'s generative
  extension that were not in the initial repertoire.}
\label{tab:phi-ablation}
\centering
\small
\resizebox{\columnwidth}{!}{%
\begin{tabular}{@{}lcccc@{}}
\toprule
\textbf{Config.} & \textbf{NERVE dims.} & \textbf{V sub-surfaces}
& \textbf{Iterations} & \textbf{Novel} \\
& \textbf{covered} & \textbf{confirmed} & & \textbf{engines} \\
\midrule
Static-EEGle & 4/5 & 3/4 & 3 & 0 \\
EEGle+$\Phi$ & 5/5 & 4/4 & 11 & 3 \\
\bottomrule
\end{tabular}%
}
\end{table}

Table~\ref{tab:phi-ablation} shows that Static-EEGle misses the
E-desynchronisation (TDA) dimension entirely---the static evasion engine
applies only standard $L_\infty$-bounded perturbations and does not attempt
temporal injection without explicit direction from~$\Phi$.  It also fails to
confirm V4 (memory-unsafe SDK), because exploiting the confused-deputy path
requires cross-referencing the System engine's format-string findings with the
Evasion engine's injection capability---a cross-dimension inference that only
$\Phi$'s stateful analysis performs.  The three LLM-synthesised engines
generated at runtime included a BLE replay-injection module, an augmented-NRA
variant targeting cross-session fingerprinting, and a supply-chain simulation
engine for E1 dependency confusion.  These would not exist in a static
repertoire.  The additional iterations (11 vs.\ 3) reflect $\Phi$ issuing
refinement plans in response to partial failures---for example, discovering
that the initial NFA parameters failed preprocessing and re-issuing with
corrected spectral envelopes.

\section{Synthetic EEG Data Generation}\label{NFA-data-gen}


\textbf{Motor Imagery.} Motor imagery (MI) involves mentally rehearsing motor actions, leading to distinct EEG patterns that reflect cortical activation and inhibition dynamics. Rather than constructing a baseline signal from parametric components, we synthesise MI trials by applying physiologically motivated frequency-band suppression directly to real resting-state epochs. This preserves the spectral structure, $1/f$ characteristics, artifacts, and cross-channel covariance of genuine neural recordings while producing class-discriminative ERD patterns indistinguishable---to a preprocessing pipeline---from real MI activity.

\paragraph{Rest Epoch as Baseline.}
Let $x_{\text{ref}} \in \mathbb{R}^{C \times T}$ denote a resting-state epoch with $C$ channels and $T$ time samples, drawn from a public dataset (for example, BCI Competition IV
2a~\cite{tangermann2012review} or
neuroTUM~\cite{tscherniak2026neurotum}). This epoch serves as the baseline signal directly:
\begin{equation}
  \text{EEG}_{\text{baseline}}(ch, t) = x_{\text{ref}}(ch, t)
  \label{eq:baseline}
\end{equation}

\paragraph{Motor Imagery Modulation.}
MI patterns are realised by subtracting band-limited components of the rest epoch at class-specific spatial locations, mimicking event-related desynchronization (ERD):
\begin{equation}
  \hat{x} = x_{\text{ref}} - S_\mu - S_\beta
  \label{eq:synth}
\end{equation}
where each suppression term is defined as:
\begin{align}
  S_\mu  &= D_\mu  \cdot \text{BP}_{[8,12]}(x_{\text{ref}}) \odot M(c^{*}) \label{eq:mu_supp} \\
  S_\beta &= D_\beta \cdot \text{BP}_{[13,30]}(x_{\text{ref}}) \odot M(c^{*})  \label{eq:beta_supp}
\end{align}
Here $\text{BP}_{[f_1, f_2]}$ denotes 4th-order zero-phase Butterworth bandpass filtering (implemented via \texttt{sosfiltfilt}), $D_\mu = 0.90$ and $D_\beta = 0.70$ are the suppression depths for the mu and beta bands respectively, $M(c^{*})$ is a class-specific spatial mask, and  $\odot$ denotes element-wise multiplication.

\paragraph{Spatial Mask.}
Suppression is centred on the focal electrode for the target class and falls off as a Gaussian over 2D Euclidean scalp distance:
\begin{equation}
  M(ch) = \exp\!\left( -\frac{d(ch,\, \text{focal}(c^{*}))^2}{2\sigma^2} \right), \quad \sigma = 0.40
  \label{eq:spatial}
\end{equation}
The focal electrode assignments follow established MI neurophysiology:

\begin{table}[h]
\centering
\begin{tabular}{lll}
\toprule
\textbf{Target class} & \textbf{Focal channel} & \textbf{Physiological basis} \\
\midrule
Left hand MI  & C4 & Contralateral (right hemisphere) ERD \\
Right hand MI & C3 & Contralateral (left hemisphere) ERD \\
Feet MI       & Cz & Central midline ERD \\
\bottomrule
\end{tabular}
\end{table}

\paragraph{Temporal Envelope.}
A sigmoid function ramps suppression in over time:
\begin{equation}
  E(t) = \frac{1}{1 + \exp\bigl(-k \cdot (t - t_{\text{onset}})\bigr)}, \quad k = 15.0,\; t_{\text{onset}} = 0.0\;\text{s}
  \label{eq:envelope}
\end{equation}

\paragraph{Complete Combination.}
The complete synthetic signal is thus:
\begin{equation}
  \hat{x} = x_{\text{ref}} - D_\mu \cdot \text{BP}_{[8,12]}(x_{\text{ref}}) \odot M(c^{*}) - D_\beta \cdot \text{BP}_{[13,30]}(x_{\text{ref}}) \odot M(c^{*})
  \label{eq:complete}
\end{equation}
No additional noise or artifact injection is required: all physiological noise, blink artifacts, and non-stationarity present in $x_{\text{ref}}$ are retained in $\hat{x}$, which is precisely what makes the synthetic signal plausible to the preprocessing pipeline. Producing a different MI class from the same rest epoch requires only changing the focal electrode in $M(c^{*})$, so arbitrarily many labelled trials can be generated on demand from a single reference recording.

\section{NRA pipeline details}
\label{app:nra}
Augmentation in the frequency space starts by transforming each EEG channel (or each epoch--channel pair) into the frequency domain via a Fast Fourier Transform (FFT). For a given channel, the complex spectrum was expressed in terms of amplitude and phase components, $A(f)$ and $\phi(f)$, respectively. To introduce controlled variability, small relative Gaussian noise was applied to the amplitudes:

\[
A'(f) = A(f) \cdot (1 + \varepsilon), \quad \varepsilon \sim \mathcal{N}(0, \sigma)
\]

where $\sigma$ denotes a small standard deviation (e.g., 0.01--0.05). Optionally, the noise magnitude can be modulated across frequency bands---for example, using a smaller $\sigma_{\text{in}}$ within discriminative ranges such as the mu or beta bands (e.g., $\sigma_{\text{in}} = 0.005$, $\sigma_{\text{out}} = 0.03$).

The perturbed signal was then reconstructed in the time domain using the inverse FFT (IFFT):

\[
x'(t) = \text{IFFT}\left( A'(f) \cdot e^{i \phi(f)} \right)
\]

The implemented function operates on an EEG epoch of shape $(3, 385)$, corresponding to three channels with 385 samples each. It performs the FFT per channel, perturbs each frequency bin’s amplitude according to the formulation above, and reconstructs the modified signal. Optionally, specific frequency bands can be preserved by applying reduced noise within those regions.

To introduce minimal temporal variability while preserving the overall oscillatory structure, a small random perturbation was applied to the phase spectrum. Specifically, the phase of each frequency component was jittered according to

\[
\phi'(f) = \phi(f) + \delta, \quad \delta \sim \mathcal{N}(0, \sigma_{\text{phase}})
\]

We add Gaussian-distributed phase noise, centered at zero, independently to each frequency bin. The standard deviation, $\sigma_{\text{phase}}$, is kept small (e.g., 0.01--0.1 radians) to prevent temporal distortion. This method is considered safer for oscillatory motor imagery paradigms where class information is mainly in the mu and beta band amplitudes, not precise phase.

To introduce controlled spectral variability, the amplitude spectrum is modulated in specific frequency bands $f \in [f_{\text{low}}, f_{\text{high}}]$. The amplitude is scaled by a small, uniformly distributed random factor $k \sim \mathcal{U}(-\alpha, \alpha)$, such that $A'(f) = A(f) \cdot (1 + k)$. For example, the 8–13 Hz mu band can be multiplied by $1+k$ to introduce mild, shape-preserving variability ($\alpha=0.03$). This function allows users to specify arbitrary ranges and scaling intervals. Unlike global noise addition, this selective modulation targets physiologically relevant bands, making it a more interpretable and neurophysiologically grounded augmentation strategy.

A conservative composite augmentation strategy was also implemented, mildly perturbing both amplitude and phase concurrently. This "combination" approach makes small adjustments to spectral magnitude and phase while preserving critical discriminative frequency bands (e.g., $\mu$ and $\beta$ rhythms) for MI classification. The modified signal is formally:

$$
x'(t) = \text{IFFT}\left( A'(f) \cdot e^{i \phi'(f)} \right),
$$

where $A'(f)$ and $\phi'(f)$ use the respective noise models with reduced perturbation magnitudes compared to independent application. This results in conservative spectral augmentation, maintaining physiological plausibility and enhancing data diversity.

\section{White-box attacks: formal definitions and algorithmic details}
\label{app:whitebox}

\subsection{Notation}
Let $\mathbf{X}\in\mathbb{R}^{d}$ denote an input, $y$ its true label, $f(\cdot)$ the model logits (or predictive function) and $L(\mathbf{X},y)$ the scalar loss used for training (e.g., cross-entropy). Let $\mathcal{T}$ denote a distribution of preprocessing transforms for EOT; $\mathbb{E}_{\tau\sim\mathcal{T}}[\cdot]$ denotes expectation over $\tau$. For a perturbation budget $\epsilon$ we write $B_\epsilon(\mathbf{X})$ for the corresponding norm ball (e.g., $L_\infty$ or $L_2$).

\subsection{Fast Gradient Sign Method (FGSM)}
A single-step $L_\infty$ update for an untargeted attack is given by
\begin{equation}
\mathbf{X}^{\text{adv}} = \mathbf{X} + \epsilon\;\mathrm{sign}\!\big(\nabla_{\mathbf{X}} L(\mathbf{X},y)\big).
\end{equation}
To reduce label leaking when the true label is not used, replace $y$ with the model prediction $\hat{y}=\arg\max f(\mathbf{X})$. For targeted FGSM, use $-\nabla_{\mathbf{X}}L(\mathbf{X},y^\star)$ where $y^\star$ is the target class.

\subsection{Projected Gradient Descent (PGD)} 
We run $T$ iterations with step size $\alpha$ and project each iterate back into $B_\epsilon(\mathbf{X})$. The attack is initialized from the clean input because random initialization was disabled: 
\begin{align} 
g_t &:= 
\nabla_{\mathbf{X}} 
L\bigl(f(\mathbf{X}^{\mathrm{adv}}_{t}),\,y\bigr), 
\\[4pt] 
\mathbf{X}^{\mathrm{adv}}_0 
&= \mathbf{X}, 
\\[4pt] 
\mathbf{X}^{\mathrm{adv}}_{t+1} 
&= 
\Pi_{B_\epsilon(\mathbf{X})} 
\!\left( 
\mathbf{X}^{\mathrm{adv}}_{t} 
+ \alpha\,\mathrm{sign}(g_t) 
\right). 
\end{align} 
Here, $\Pi_{B_\epsilon(\mathbf{X})}(\cdot)$ denotes projection onto the feasible perturbation ball. For $L_2$ variants, the sign update is replaced by a normalized gradient direction. We stop early when the example becomes misclassified.

\subsection{Carlini \& Wagner (C\&W)}
C\&W formulates an unconstrained optimization that trades perturbation norm against an attack loss. For the $L_2$ formulation used here:
\begin{equation}
\min_{\delta}\; \|\delta\|_2^2 \;+\; c\;\mathbb{E}_{\tau\sim\mathcal{T}}\big[ g\big(f(\tau(\mathbf{X}+\delta)), y\big)\big],
\end{equation}
where $g(\cdot)$ is a differentiable misclassification objective (e.g., margin-based) and $c>0$ is a weighting constant. Optimization proceeds with Adam and optional restarts over $c$ to identify a minimal perturbation achieving success. The perturbation is constrained implicitly by a penalty or via an explicit projection step, depending on the chosen implementation.

\subsection{DeepFool}
Assuming a differentiable classifier with decision boundaries locally approximated by linear functions, DeepFool iteratively computes minimal perturbations to cross the closest boundary. At iteration $i$:
\begin{align}
\delta_i &= -\frac{f(\mathbf{X}_i)}{\|\nabla f(\mathbf{X}_i)\|_2^2}\,\nabla f(\mathbf{X}_i),\\[4pt]
\mathbf{X}_{i+1} &= \mathbf{X}_i + \delta_i,
\end{align}
and the procedure repeats until the predicted label changes. The total perturbation is $\delta=\sum_i \delta_i$, which serves as an estimate of the minimal $L_2$ perturbation required.

\subsection{Implementation notes} 
\begin{itemize} 
    \item All attacks were executed using the same preprocessing pipeline to ensure fair comparison. EOT was not enabled for the white-box robustness probe.
    \item For iterative attacks, we used early stopping on misclassification and report perturbation norms measured on the final clipped example.
    \item Attack budgets and iteration counts are given in the caption of Table~\ref{tab:art-attack-comparison}. PGD used a step size of $\alpha=0.1$ (\texttt{eps\_step} in ART). For C\&W, the initial optimization constant was $c=0.01$ (\texttt{initial\_const}), with \texttt{binary\_search\_steps}=10. No explicit random seed was set, and PGD random initialization was disabled. 
\end{itemize}

\section{Cross-Layer Attack Chain: V$\to$N End-to-End}
\label{app:vn-chain}

The V and N dimensions are not independent: Vein Tapping provides the
injection channel through which NFA payloads reach the target model.
To demonstrate this cross-layer dependency explicitly, we performed an
end-to-end V$\to$N chain on NeuroSky MindWave Mobile~2.

\textbf{Setup.}
Using the V1 passive-sniffing configuration (nRF52840 dongle, Wireshark
GATT plugin), we first established the BLE advertisement timing and GATT
handle layout of the target device.  We then deployed the V2 MitM
configuration (two Raspberry Pi~4 nodes running GATTacker), establishing
a transparent relay between the headset and the host application.  From
this relay position we can observe, modify, or replace any epoch before it
reaches the BrainFlow ingestion layer.

\textbf{Injection.}
NFA payloads (A1--A3 from Table~\ref{tab:17attacks}) were pre-generated
offline and serialised to the raw byte layout expected by the NeuroSky
TGAM packet format (24-bit attention/meditation values with checksum).
The relay intercepts the real EEG packet, discards it, and injects the
NFA-synthesised payload within the same BLE connection interval
($\le$7.5\,ms), making the substitution transparent to the host.

\textbf{Result.}
The BrainFlow mental-state classifier received the injected epochs and
produced the attacker-chosen class output in 100\,\% of injection attempts
across 20 trials, with a mean end-to-end latency from interception to
misclassification of 9.3\,ms (dominated by BrainFlow's preprocessing
pipeline).  The host application observed no anomaly: packet timing,
checksum, and attention-value range were all within normal bounds.
This confirms that the V--N attack chain is not a theoretical composition
but an empirically executable exploit requiring only commodity hardware and
the NFA payload generation capability provided by EEGle.

\section{Experimental Setup}
\label{app:setup}
Table~\ref{tab:sysconfig} lists the workstation, software, and BCI hardware used
for every experiment reported in Section~\ref{sec:evaluation}.  The mapping from
each platform to the \NERVE dimension it was used to evaluate is given in
Table~\ref{tab:eval-platform}.

\begin{table}[h!]
\centering
\caption{System Configuration}
\label{tab:sysconfig}
\renewcommand{\arraystretch}{1.2} 
\begin{tabular}{|p{2.4cm}|p{5.3cm}|}
\hline
\textbf{Category} & \textbf{Details} \\
\hline
Workstation & Alienware Aurora R16 \\
\hline
Operating System & Ubuntu 25.04 (Linux kernel 6.14.0-34) \\
\hline
Hardware & NVIDIA GeForce RTX 4070 Ti GPU; 64 GB RAM \\
\hline
Programming Environment & Python 3.13.3 \\
\hline
Core Libraries & TensorFlow 2.20.0; NumPy 2.3.2; SciPy 1.16.1; scikit-learn 1.7.1; Pandas 2.3.2; Matplotlib 3.10.6; MNE 1.10.1 \\
\hline
BCI Tools & BrainFlow Library 5.18.0 \\
\hline
Devices & OpenBCI Cyton (8-channel, 32-bit); NeuroSky MindWave Mobile 2; Muse2 \\
\hline
\end{tabular}
\end{table}

\section{Model Training Hyperparameters}
\label{app:hyperparameters}
The hyperparameters for the EEGNet base model and for the transfer learning setup used in the backdoored multi-trigger model are summarised in Figure~\ref{fig:eegnet-hparams} and Figure~\ref{fig:bd-transfer-hparams}, respectively.

\begin{figure*}[h!]
\centering
\begin{minipage}[t]{0.48\textwidth}
\small
\renewcommand{\arraystretch}{1.5}

\begin{tabular}{|p{4.5cm}|p{3.2cm}|p{5.2cm}|}
\hline
\textbf{Category} & \textbf{Hyperparameter} & \textbf{Value} \\ \hline

\multirow{3}{*}{Data Sampling \& Segmentation}
    & Sampling Rate & 160 Hz \\ \cline{2-3}
    & Epoch Length & 3 seconds \\ \cline{2-3}
    & Segment Length & 480 samples \\ \hline

Data Split & Train / Val / Test & 70\% / 15\% / 15\% \\ \hline

\multirow{2}{*}{Preprocessing}
    & Band-pass Filter & 1--40 Hz (IIR) \\ \cline{2-3}
    & Normalization & StandardScaler (per-segment, channel-wise) \\ \hline

\multirow{8}{*}{Model Architecture (EEGNet)}
    & Channels & 64 \\ \cline{2-3}
    & Samples per Epoch & 480 \\ \cline{2-3}
    & Temporal Kernel Length & 64 \\ \cline{2-3}
    & $F_1$ (Temporal Filters) & 8 \\ \cline{2-3}
    & Depth Multiplier ($D$) & 2 \\ \cline{2-3}
    & $F_2$ (Separable Filters) & 16 \\ \cline{2-3}
    & Dropout Rate & 0.5 \\ \cline{2-3}
    & Activation Function & ELU \\ \hline

\multirow{5}{*}{Training Parameters}
    & Optimizer & Adam \\ \cline{2-3}
    & Learning Rate & 0.001 \\ \cline{2-3}
    & Loss Function & Categorical Cross-Entropy \\ \cline{2-3}
    & Batch Size & 32 \\ \cline{2-3}
    & Epochs & 100 (early stopping) \\ \hline

\multirow{3}{*}{Callbacks}
    & EarlyStopping & Patience 15, monitor: val\_accuracy \\ \cline{2-3}
    & Reduce LR on Plateau & Factor 0.2, Patience 3, Min LR $=10^{-7}$ \\ \cline{2-3}
    & Model Checkpoint & Monitor: val\_accuracy \\ \hline

\end{tabular}

\caption{EEGNet Base Model Hyperparameters}
\label{fig:eegnet-hparams}
\vspace{1em}

\begin{tabular}{|p{4.5cm}|p{3.2cm}|p{5.2cm}|}
\hline
\textbf{Category} & \textbf{Hyperparameter} & \textbf{Value} \\ \hline

\multirow{2}{*}{Data Sampling \& Segmentation}
    & Sampling Rate & 160 Hz \\ \cline{2-3}
    & Epoch Length & 3 seconds \\ \hline

Data Split & Train / Val / Test & 70\% / 15\% / 15\% \\ \hline

\multirow{2}{*}{Preprocessing}
    & Band-pass Filter & 1--40 Hz (IIR) \\ \cline{2-3}
    & Normalization & None (TL stage) \\ \hline

\multirow{2}{*}{Backdoor Injection}
    & Injection Strategy & Added to all epochs of class \\ \cline{2-3}
    & Channel Application & All channels of poisoned samples \\ \hline

Transfer Learning Setup & Trainable Layers & All unfrozen \\ \hline

\multirow{5}{*}{Training Parameters}
    & Optimizer & Adam \\ \cline{2-3}
    & Learning Rate & 0.0005 \\ \cline{2-3}
    & Loss Function & Categorical Cross-Entropy \\ \cline{2-3}
    & Batch Size & 32 \\ \cline{2-3}
    & Epochs & 100 (early stopping) \\ \hline

\multirow{2}{*}{Callbacks}
    & EarlyStopping & Patience 10, monitor: val\_accuracy \\ \cline{2-3}
    & Reduce LR on Plateau & Factor 0.2, Patience 5, Min LR $=10^{-7}$ \\ \hline

\end{tabular}

\caption{Transfer Learning Hyperparameters for Backdoored Multi-Trigger Model}
\label{fig:bd-transfer-hparams}
\end{minipage}
\hfill
\begin{minipage}[t]{0.48\textwidth}
\end{minipage}

\end{figure*}

\section{Backdoor Figures}
\begin{figure*}[t]
    \centering
    
    \begin{subfigure}{0.46\linewidth}
        \centering
        \includegraphics[width=\linewidth]{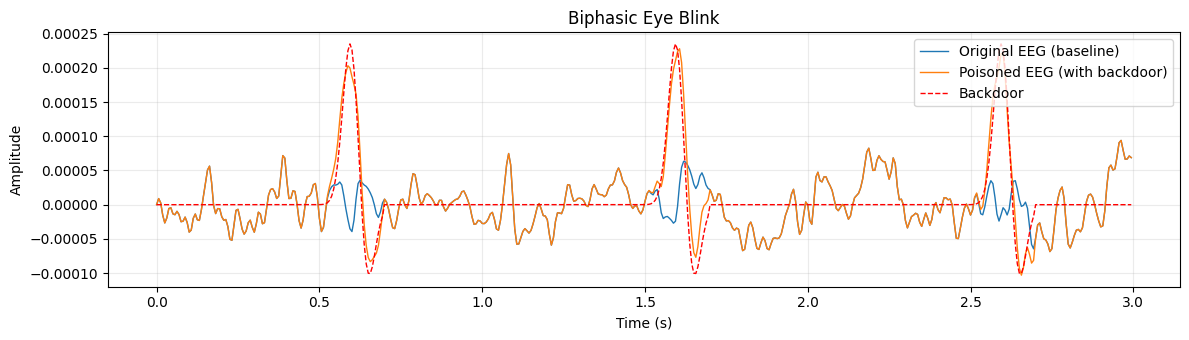}
        \caption{Biphasic Eye Blink - Top: original EEG; Middle: poisoned EEG with injected blink template; Bottom: Biphasic eye-blink backdoor pattern.}
        \label{fig:bd-biphasic}
    \end{subfigure}\hfill
    \begin{subfigure}{0.46\linewidth}
        \centering
        \includegraphics[width=\linewidth]{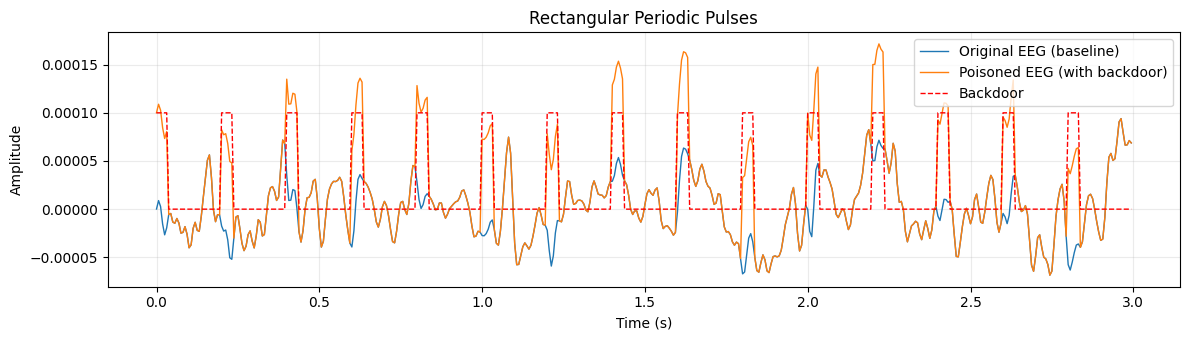}
        \caption{Rectangular Periodic Pulse - clean vs. poisoned EEG and backdoor waveform.}
        \label{fig:bd-npp}
    \end{subfigure}\\[0.8em]

    \begin{subfigure}{0.46\linewidth}
        \centering
        \includegraphics[width=\linewidth]{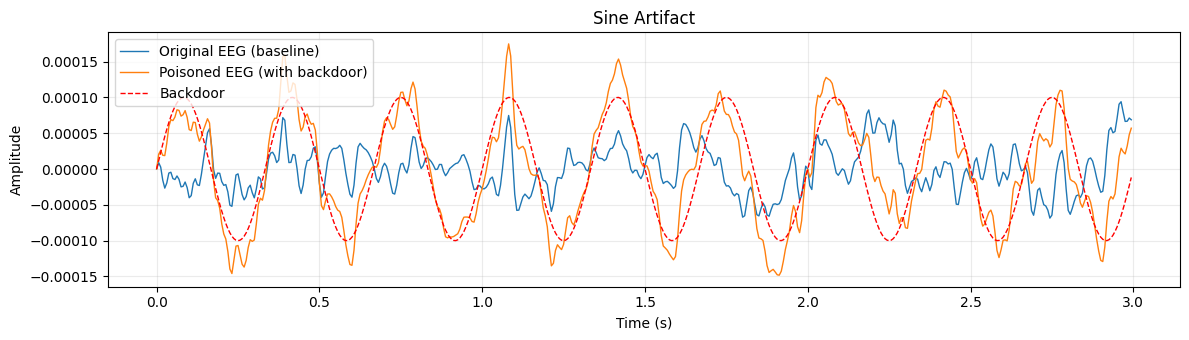}
        \caption{Continuous Sine - original, poisoned, and backdoor waveform (3~Hz sine).}
        \label{fig:bd-sine}
    \end{subfigure}\hfill
    \begin{subfigure}{0.46\linewidth}
        \centering
        \includegraphics[width=\linewidth]{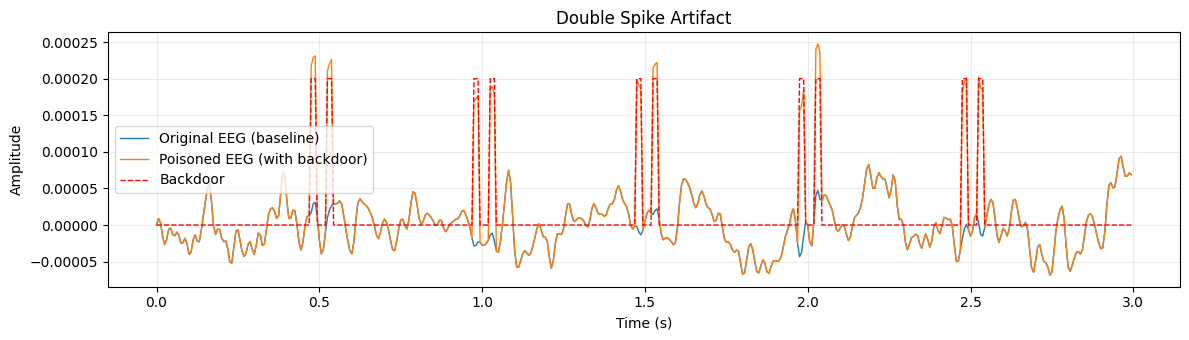}
        \caption{Double Spike - original vs. poisoned EEG and injected double-spike backdoor.}
        \label{fig:bd-double}
    \end{subfigure}\\[0.8em]

    \begin{subfigure}{0.46\linewidth}
        \centering
        \includegraphics[width=\linewidth]{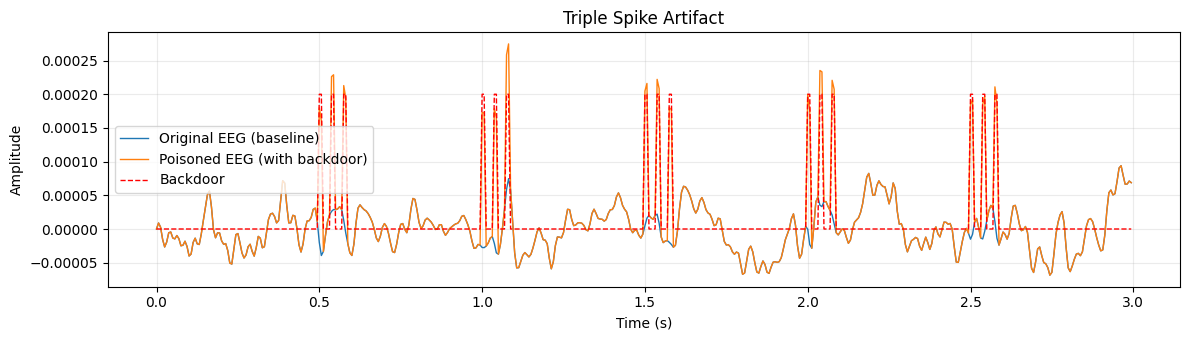}
        \caption{Triple Spike - clean vs. poisoned EEG and backdoor waveform.}
        \label{fig:bd-triple}
    \end{subfigure}\hfill
    \begin{subfigure}{0.46\linewidth}
        \centering
        \includegraphics[width=\linewidth]{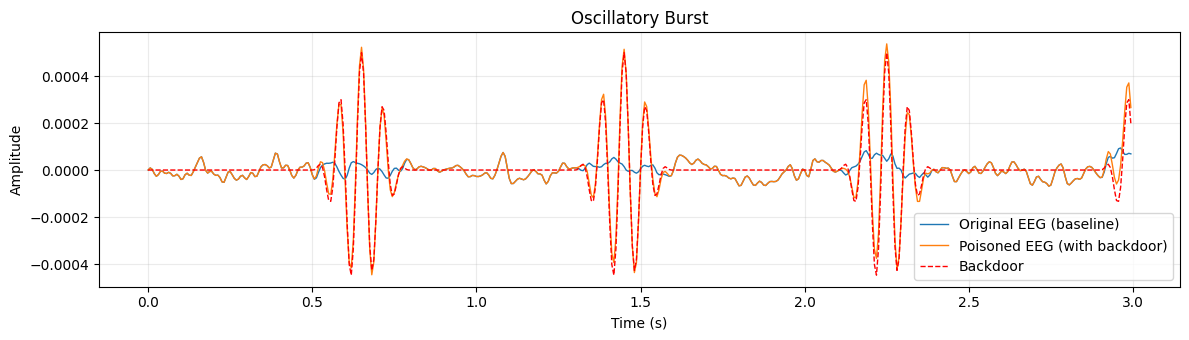}
        \caption{Oscillatory Burst - original, poisoned, and backdoor waveform (15~Hz rhythmic bursts).}
        \label{fig:bd-oscillatory}
    \end{subfigure}\\[0.8em]

    \begin{subfigure}{0.46\linewidth}
        \centering
        \includegraphics[width=\linewidth]{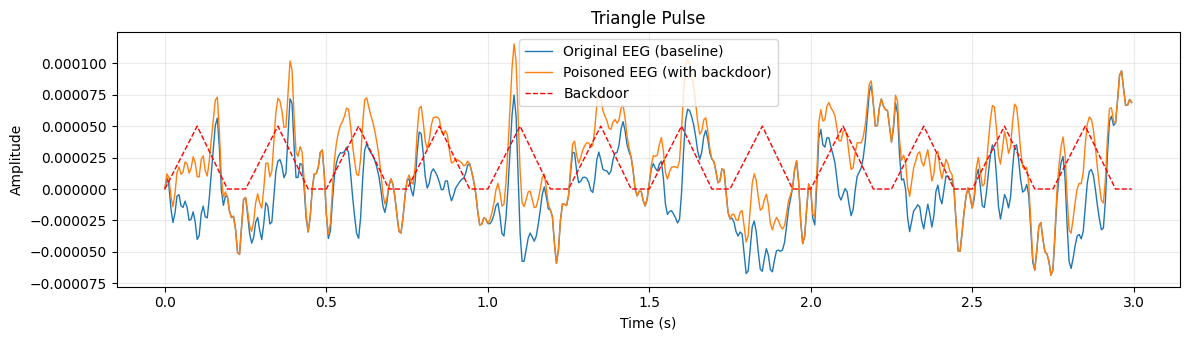}
        \caption{Triangle Pulse - original vs. poisoned EEG and backdoor triangular pulses.}
        \label{fig:bd-triangle}
    \end{subfigure}\hfill
    \begin{subfigure}{0.46\linewidth}
        \centering
        \includegraphics[width=\linewidth]{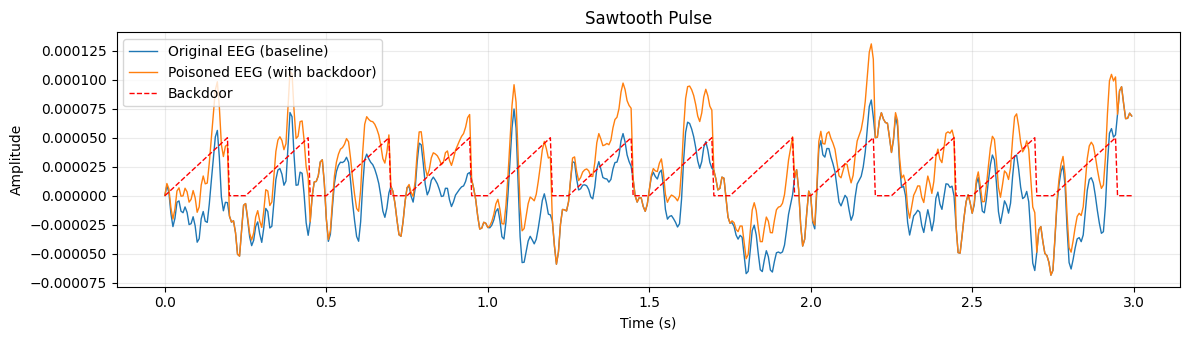}
        \caption{Sawtooth Pulse - original vs. poisoned EEG and injected sawtooth waveform.}
        \label{fig:bd-sawtooth}
    \end{subfigure}\\[0.8em]

    \begin{subfigure}{0.46\linewidth}
        \centering
        \includegraphics[width=\linewidth]{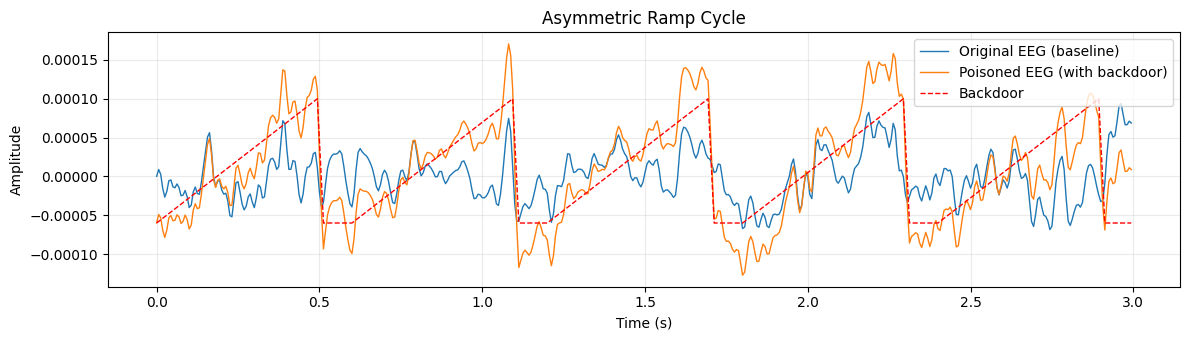}
        \caption{Asymmetric Ramp Cycle - original vs. poisoned EEG and injected ramp waveform.}
        \label{fig:bd-ramp}
    \end{subfigure}\hfill
    \begin{subfigure}{0.46\linewidth}
        \centering
        \includegraphics[width=\linewidth]{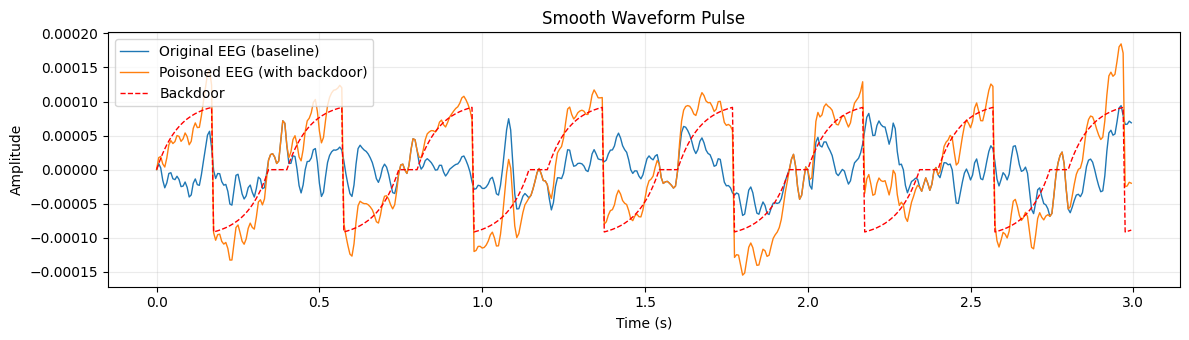}
        \caption{Smooth Waveform Pulse - original vs. poisoned EEG and injected smooth waveform pattern.}
        \label{fig:bd-wave}
    \end{subfigure}

    \caption{Overview of backdoor waveform templates injected into EEG signals across multiple patterns.}
    \label{fig:bd-all}
\end{figure*}

Figure~\ref{fig:bd-all} summarises the ten backdoor waveform templates we consider, illustrating how each pattern is injected into clean EEG to produce the poisoned signals used in our experiments.

\end{document}